\documentclass[10pt,a4paper,twocolumn,english,floatfix,groupedaddress,superscriptaddress,showpacs]{revtex4}

\usepackage{graphicx}
\usepackage{epsfig}
\usepackage[english]{babel}
\usepackage{amsmath}
\usepackage{amssymb}
\usepackage{amsfonts}
\usepackage{longtable}
\usepackage{dcolumn}
\usepackage{bm}
\usepackage{bbm}

\begin{document}
\title{Optimized pulses for the perturbative decoupling of
spin and decoherence bath}

\author{S. Pasini}
\affiliation{Lehrstuhl f\"{u}r Theoretische Physik I, 
Technische Universit\"{a}t Dortmund,
 Otto-Hahn Stra\ss{}e 4, 44221 Dortmund, Germany}

\author{P. Karbach}
\affiliation{Lehrstuhl f\"{u}r Theoretische Physik I, 
Technische Universit\"{a}t Dortmund,
 Otto-Hahn Stra\ss{}e 4, 44221 Dortmund, Germany}

\author{C. Raas}
\affiliation{Lehrstuhl f\"{u}r Theoretische Physik I, 
Technische Universit\"{a}t Dortmund,
 Otto-Hahn Stra\ss{}e 4, 44221 Dortmund, Germany}

\author{G. S. Uhrig}
\affiliation{Lehrstuhl f\"{u}r Theoretische Physik I, 
Technische Universit\"{a}t Dortmund,
 Otto-Hahn Stra\ss{}e 4, 44221 Dortmund, Germany}

\date{\textrm{\today}}

\begin{abstract} 
In the framework of nuclear magnetic resonance (NMR), 
we consider the general problem of the 
coherent control of a spin coupled to a bath  by means of composite or 
continuous pulses of duration $\tau_\mathrm{p}$. We show explicity
that it is possible to design the  pulse in order to achieve a decoupling 
of the spin from the bath up to the third order in $\tau_\mathrm{p}$. 
The evolution of the system is separated in the evolution
of the spin under the action of the pulse and of the bath, 
times correction terms. We derive the  correction terms for a general 
time dependent  axis of rotation and for a general coupling 
between the spin and the environment. The resulting corrections
can be made vanish by an appropriate design of the pulse.
For $\pi$ and $\pi/2$ pulses, we demonstrate explicitly
that pulses exist which annihilate the first and the second order corrections
even if the bath is fully quantum mechanical, i.e., it displays
 internal dynamics. Such pulses will also be useful for quantum
information processing.
\end{abstract}


\maketitle

\section{Introduction}\label{Intro}

The decay of the spin polarization and  the occurrence of 
systematic errors due to the coupling to the environment
has always been one of the main difficulties to overcome in 
high-precision experiments in nuclear magnetic resonance (NMR)
and in magnetic resonance imaging (MRI).
Hence it is a long standing issue to reduce the influence
of the coupling to the environment. One way to achieve this goal
is dynamic control, i.e., the application of suitable control pulses. 

The very first step in this direction was the done
by Hahn in 1950 who observed that a $\pi$ pulse
after the delay time $\tau$ evokes a spin echo at 
$2\tau$ \cite{hahn50}. Further developments comprise
the iteration of the two-pulse cycle 
$\tau-\pi-2\tau-\pi-\tau$ according to Carr, Purcell, Meiboom, 
and Gill \cite{carr54,meibo58} and more sophisticated sequences in NMR
\cite{haebe76,ernst87,freem98}.
In quantum information processing (QIP) this
 kind of approach was found and used
under the name of dynamic decoupling (DD)
\cite{viola98,ban98,viola99a,khodj05}. A vital point for QIP
is to consider open quantum systems which induce the decoherence
and hence the loss of information.
In particular, the
instants in time, at which the $\pi$ pulses 
are applied, can be optimized \cite{uhrig07,lee08a,yang08}
which has been verified very recently in experiment \cite{bierc09a}.

Besides the optimization of the sequence, the individual
pulse can be designed to fit its purpose best. Theoretically,
the instantaneous $\delta$ peak pulse is the optimum choice.
But it cannot be realized experimentally. Hence the design
of the real pulses matters very much in practice. Again, this was
first seen in NMR where composite pulses
are discussed extensively \cite{tycko83,levit86}
and their importance for quantum computation
is recognized \cite{cummi00,cummi03,fortu02,brown04,motto06,alway07}.
An efficient numerical technique to tailor pulses employed
in NMR is optimal control theory \cite{skinn03,kobza04,luy05,gersh07,gersh08}.
The main aim is to find pulses which are robust against static
resonance offsets and miscalibrations of the pulses.
Sengupta and Pryadko suggested soft pulses, i.e.,
pulses of continuous shape so that their frequency
selectivity is better, in order to mitigate
the coupling to other parts of the system or to
the environment \cite{sengu05}. The proposed pulses
make the second order corrections  to zero if these corrections
result from a static perturbation, for instance a resonance offset
\cite{pryad08a,pryad08b}.

In previous work, we investigated a similar issue 
\cite{pasin08a,karba08,pasin08b}. The aim
was to disentangle the pulse dynamics from the dynamics
of the system which comprised spin, bath and the coupling
between them. This aim is the natural one if the pulse shall be used
as drop-in for an instantaneous pulse in a DD sequence. Before and after the
instantaneous $\delta$ spike the complete 
system including the spin-bath coupling
 is dynamic in the DD sequence and the same
must be true for the drop-in, unless
the finite duration of the experimental pulse is 
accounted for otherwise \cite{bierc09b,uhrig09c}. This means that
the spin-bath coupling may not be set to zero in
the equivalent description of the real pulse of
duration $\tau_\mathrm{p}$. Interestingly, for the
disentanglement ansatz, we found that $\pi$
pulses can have vanishing linear corrections, but it
is rigorously impossible that their second order
correction vanishes: this is the no-go theorem for the disentanglement
of pulse and system \cite{pasin08a,pasin08b}.

In the present work we show by explicit construction
that the no-go theorem does not apply if we aim at
averaging the coupling between spin and bath to
zero during the duration of the pulse. This is the
aim pursued by many preceding studies 
\cite{tycko83,levit86,cummi00,cummi03,fortu02,brown04,sengu05,motto06,alway07,skinn03,kobza04,luy05,gersh07,gersh08,pryad08a,pryad08b}.
Our particular achievements are twofold. First, we derive
the corrections for general pulse shape (arbitrary time dependent
axis of rotation, arbitrary amplitude) and for general coupling
between spin and a bath with all possible
quantum fluctuations. Second, we find solutions for $\pi$ and
$\pi/2$ pulses which are correct in all second order corrections,
even for a dynamical bath. Thereby, our results go significantly 
beyond previous findings.

After this general introduction, the model and the motivation for our approach 
are discussed in Sect.\ \ref{MM} while the technique is developed in Sects.\ 
\ref{GenEq} and \ref{Expansion}. The general equations are  presented 
in Sect.\ \ref{Discussion} for a time dependent axis of rotation and a general
 coupling. In Sect.\ \ref{Solutions} we solve the equations for the specific 
case  of a fixed axis of rotation and a coupling along the $z$ axis. 
Finally, we conclude our study in Sect.\ \ref{Conclusions}.

\section{Model and motivation}
\label{MM}
We consider in the beginning the most general case of a single spin coupled 
to a bath 
\begin{equation}
 \label{Hamilt_bq}
H=H_\mathrm{b}+\vec{\sigma}\cdot\vec{A} ,
\end{equation} 
where $\vec\sigma$ is the vector of Pauli matrices. This is the 
most general case because no spin direction is singled out. The generality of 
this Hamiltonian comprises the simple case of a spin coupled to the bath only 
along the $z$ direction. This corresponds to the common limit where the 
longitudinal relaxation time $T_1$ is much longer than the transverse 
relaxation time $T_2$. The internal energy scale of the bath $H_\mathrm{b}$ 
shall be denoted with $\omega_\mathrm{b}=||H_\mathrm{b}||$, while 
$\lambda=||\vec{A}||$  represents the average strength of 
the coupling between the spin and the bath.

The Hamiltonian of the control pulse reads
\begin{equation}
 \label{Hamilt_pulse}
H_0(t)=\vec{\sigma}\cdot\vec{v}(t) ,
\end{equation} 
where $\vec{v}(t)=(v_x(t),v_y(t),v_z(t))$ is a time dependent vector defining 
the shape of the pulse. The axis of rotation at the instant $t$ is given by 
the unit vector $\vec{v}(t)/|\vec{v}(t)|$.

We are interested in studying how the system evolves between 0 and 
$\tau_\mathrm{p}$,  which is the duration time  of the pulse. 
The total Hamiltonian is  $H_\mathrm{tot}=H+H_0(t)$. The system, built from 
the spin and the bath, evolves according to the evolution operator
\begin{equation}
 \label{EvOp_gen}
U_\mathrm{p}(\tau_\mathrm{p},0)=T\left[e^{-\mathrm{i}H\tau_\mathrm{p}-
\mathrm{i}\int_0^{\tau_\mathrm{p}}H_0(t)\mathrm{d}t}\right], 
\end{equation} 
where $T$ stands for the standard time ordering.

In general, we denote the rotation of the spin due to 
the pulse $\vec{v}(t)$  by $\hat{P}_\theta$. In previous papers 
we approximated  the time evolution operator according to the ansatz
\begin{equation}
\label{eq:ansatz-dd}
 U_\mathrm{p}(\tau_\mathrm{p},0)\approx 
e^{-\mathrm{i}(\tau_\mathrm{p}-\tau_\mathrm{s})H}
\hat{P}_\theta e^{-\mathrm{i}\tau_\mathrm{s}H}, 
\end{equation} 
where $\tau_\mathrm{s}\in[0,\tau_\mathrm{p}]$ represents the instant at which 
the idealized $\delta$ peak must be situated to
approximate the real pulse in leading order. As mentioned
already in the
introduction the ansatz \eqref{eq:ansatz-dd} is the
natural one for a \mbox{drop-in} in a DD sequence because in such a
sequence  the total Hamiltonian $H$ is active
before and after the application of the ideal, instantaneous pulse.

In many studies, the goal is the approximation
\begin{equation}
\label{goal_others}
 U_\mathrm{p}(\tau_\mathrm{p},0)\approx \hat{P}_\theta .
\end{equation} 
This is the case of CORPSE and SCORPSE pulses \cite{cummi00,cummi03}.
For $\theta=\pi$, $\tau_\mathrm{s}=\tau_\mathrm{p}/2$,
and an axis of rotation in the $xy$ plane 
the two approximations coincide as long as $H=\sum_j 
\lambda_j \sigma_z^{(j)}$ (as an exception, we here consider an arbitrary 
number of spins) due to
\begin{subequations}
\begin{eqnarray}
U_\mathrm{p}(\tau_\mathrm{p},0) &=&  e^{-\mathrm{i}(\tau_\mathrm{p}/2)H} 
\hat P_\pi e^{-\mathrm{i}(\tau_\mathrm{p}/2) H}
\quad
\\
&=& \hat P_\pi e^{\mathrm{i}(\tau_\mathrm{p}/2)H} 
e^{-\mathrm{i}(\tau_\mathrm{p}/2) H}
\\
&=& \hat P_\pi .
\end{eqnarray}
\end{subequations}

In the present work we generalize \eqref{eq:ansatz-dd}
and \eqref{goal_others} to the approximation
\begin{equation}
\label{eq:present-goal}
 U_\mathrm{p}(\tau_\mathrm{p},0)\approx e^{-\mathrm{i}\tau_\mathrm{p} 
H_\mathrm{b}}\hat{P}_\theta
\end{equation} 
where $[\hat{P}_\theta,H_\mathrm{b}]=0$ holds by definition. Hence it
does not make sense to define an instant $\tau_\mathrm{s}$ at which
the equivalent $\delta$ pulse is located. In other words,
all values $\tau_\mathrm{s}\in [0,\tau_\mathrm{p}]$ are equivalent.
The difference to the ansatz \eqref{eq:ansatz-dd}
is that the coupling between spin and bath is averaged to zero
by the pulse. The difference to the ansatz \eqref{goal_others}
is that we keep the dynamics of the bath. The latter point
is relevant only if the bath is not completely static.

Pulses which realize the approximation \eqref{eq:present-goal}
are useful in NMR for the preparation of particular states
and the subsequent measurement of the signal decay without delay
\cite{gersh07,gersh08}. In QIP, pulses of the type \eqref{eq:present-goal}
are relevant for single quantum bit gates. 
For dynamic decoupling they can also be used if the pulse
sequence takes the finite pulse duration into account \cite{uhrig09c}.

\section{General Equations}
\label{GenEq}
For the total time evolution we start from
\begin{equation}
\label{ansatz0}
U_\mathrm{p}\left(\tau_\mathrm{p} ,0\right)=
e^{-\mathrm{i}\tau_\mathrm{p}H_\mathrm{b}} \
T\left\{e^{-\mathrm{i}\vec{\sigma}\cdot \int_{0}^{\tau_\mathrm{p}}
 \vec{v}(t) \mathrm{d}t}\right\} \  {U}\left(\tau_\mathrm{p} ,0\right) .
\end{equation} 
This ansatz is consistent with the general goal \eqref{eq:present-goal}. 
The unitary $ {U}\left(\tau_\mathrm{p} ,0\right)$ incorporates the corrections 
implied by the choice of the ansatz. These corrections will be functions of 
the  coupling constants of the Hamiltonian and of the pulse shape. 
Obviously, no corrections occur if the coupling vanishes ($\lambda=0$)
so that $ {U}\left(\tau_\mathrm{p} ,0\right)$ is the identity operator. 
For the general case $\lambda\neq 0$, we search
for the conditions under which the corrections vanish.
To this end,  $ {U}\left(\tau_\mathrm{p} ,0\right)$ must be determined 
from the Schr\"odinger equation. The prerequisite is to
know the time dependence of the pulse.

The time ordered exponential in (\ref{ansatz0}) represents the
actual pulse. We describe its time dependence by
\begin{eqnarray}
 \label{Ptau}
\hat{P}_\tau &:=& T\left\{e^{-\mathrm{i}\vec{\sigma}\cdot
\int_{0}^{\tau}\vec{v}(t)\mathrm{d}t}\right\},\ \forall \tau
\\
&=& e^{-\mathrm{i}\vec{\sigma}\cdot\hat{a}(\tau){\psi(\tau)}/2} . 
\end{eqnarray} 
This expression represents an
overall rotation around the axis ${\hat a}(\tau)$ 
($|{\hat a}(\tau)|=1$) and about the angle $\psi(\tau)$.
For $\tau=\tau_\mathrm{p}$ the pulse is completed and by
definition we have $\psi(\tau_\mathrm{p})=\theta$.

The pulse satisfies the Schr\"odinger equation
\begin{equation}
 \label{Schr_P}
\mathrm{i}\partial_\tau \hat{P}_\tau=H_0(\tau)\hat{P}_\tau\,
\end{equation}
for all  $\tau$ in $[0,\tau_\mathrm{p}]$.
The relation between $\vec{v}(t)$, $\hat{a}(t)$ 
and $\psi(t)$ can be derived by solving Eq.\ (\ref{Schr_P}). 
We refer the reader to  Ref.\ \onlinecite{pasin08b} for further details.
The vector $\vec{v}(t)$ can be written  as a function of the axis and the 
angle of rotation
\begin{eqnarray}
\label{a_psi_vs_v_noSigma}\nonumber
2\vec{v}(t)&=&\psi^\prime(t)\hat{a}(t)+\hat{a}^\prime(t)\sin\psi(t)
\\
&-&(1-\cos\psi(t))\left(\hat{a}^\prime(t)\times\hat{a}(t)\right).
\end{eqnarray} 
Multiplication with $\hat{a}(t)$ yields the derivative of $\psi(t)$
\begin{equation}
\label{psi_vs_v_a}
\vec{v}(t)\cdot\hat{a}(t)={\psi^\prime(t)}/{2}.
\end{equation}
Eq.\ (\ref{a_psi_vs_v_noSigma}) can also be used to
find $\psi(t)$ and $\hat a(t)$ from  $\vec{v}(t)$ by integration.
This is the way one has to take from an experimentally given pulse 
to its theoretical description.

The Schr\"odinger equation for the total time evolution reads
\begin{equation}
\label{Schr_U_gen}
\mathrm{i}\partial_\tau U_\mathrm{p}(\tau,0)=(H+H_0(\tau))U_\mathrm{p}(\tau,0).
\end{equation} 
Together with (\ref{Schr_P}),  the following differential equation
ensues
\begin{equation}
 \label{diffEq_Utilde}
i\partial_\tau  {U}(\tau,0)=G(\tau) {U}(\tau,0),
\end{equation} 
where
\begin{equation}
 \label{function_G}
G(\tau):=e^{\mathrm{i}H_\mathrm{b}\tau}\hat{P}^{-1}_\tau
\left(\vec{\sigma}\cdot\vec{A}\right) \hat{P}_\tau 
e^{-\mathrm{i}H_\mathrm{b}\tau}.
\end{equation} 
From its definition one sees that $G(\tau)$ is always linear 
in $\vec{A}$ and thus in $\lambda$. So it vanishes if the coupling vanishes.
Note that $G(\tau)$ differs from the corresponding
time dependent operator $F(\tau)$ in Refs.\ \onlinecite{pasin08a} and 
\onlinecite{pasin08b} where $\vec{v}(\tau)$ appeared. So the present result
is to some extent simpler than the one for the approximation 
\eqref{eq:ansatz-dd}.

At $\tau=\tau_\mathrm{p}$ we formally obtain for the unitary correction
\begin{equation}
 {U}(\tau_\mathrm{p},0)=
T\left\{ e^{-\mathrm{i}\int_{0}^{\tau_\mathrm{p}}  G(t) \mathrm{d}t}\right\}.
\label{U_G}
\end{equation}
Further progress requires the explicit form of $G(t)$
\begin{equation}
 \label{Htilde}G(t)=e^{\mathrm{i}H_\mathrm{b} t}\  {H}_\mathrm{qb}\ 
e^{-\mathrm{i}H_\mathrm{b} t},
\end{equation} 
with
\begin{subequations}
\begin{eqnarray}
\label{Htilde1} 
{H}_\mathrm{qb}&:=& \hat{P}^{-1}_t\left(\vec{\sigma}\cdot\vec{A}\right) 
\hat{P}_t
\\
&=&\left[\cos\psi(t)(\vec{\sigma}\cdot\vec{A})-
\sin\psi(t)\vec{\sigma}\cdot\left(\hat{a}(t)\times\vec{A}\right)\right.
\nonumber
\\
\label{Htilde3}
&&\left.+
2\sin^2\frac{\psi(t)}{2}(\hat{a}(t)\cdot\vec{A})(\vec{\sigma}\cdot\hat{a}(t))
\right].
\end{eqnarray}
\end{subequations}  
To derive relation (\ref{Htilde3}) the following 
identities are useful
\begin{subequations}
\begin{eqnarray}
 \label{P_trig}
\hat{P}_t &=& \cos(\psi/2)-\mathrm{i}
(\vec{\sigma}\cdot\hat{a})\sin(\psi/2)\qquad
\\
(\vec{m}\cdot\vec{\sigma})(\vec{n}\cdot\vec{\sigma})&=&\vec{m}\cdot\vec{n}+
\mathrm{i}\vec{\sigma}\cdot(\vec{m}\times\vec{n})
\\
\vec{m}\times(\vec{n}\times\vec{l}) & =& \vec{n}(\vec{m}\cdot\vec{l})-
\vec{l}(\vec{m}\cdot\vec{n}),
\end{eqnarray}
\end{subequations}
where the time dependencies of  $\hat{a}(t)$ and $\psi(t)$ are omitted
to lighten the notation.

We again use the notation introduced previously \cite{pasin08b} and define
\begin{equation}
\label{S_definition}
\vec{S}(t) := \hat{P}^{-1}_t\vec{\sigma}\hat{P}_t.
\end{equation}
This vector operator represents a rotation of $\vec{\sigma}$ about the
axis $\hat{a}$ by the angle $\psi$. Hence it can also be written
as
\begin{equation}
 \label{S_rotation_matrix}
\vec{S}(t)=D_{{\hat a}}(\psi)\vec\sigma,
\end{equation} 
where $D_{{\hat a}}(\psi)$ is the $3\times 3$ dimensional
matrix describing the rotation about the axis $\hat{a}$ by the angle $\psi$. 
 This leads to
\begin{subequations}
\begin{eqnarray}
 \label{Hqb_vec_n} {H}_\mathrm{qb}=
\hat{P}_t^{-1}\left(\vec{\sigma}\cdot\vec{A}\right)\hat{P}_t
&:=&\vec{S}(t)\cdot\vec{A}\\
&=& \left(D_{\hat{a}}(\psi)\vec{\sigma}\right)\cdot\vec{A}\\
&=&\left(D_{\hat{a}}(-\psi)\vec{A}\right)\cdot\vec{\sigma}
\label{Hqb_vec_D}\\
&:=&\vec{n}_A(t)\cdot\vec{\sigma}
\label{Hqb_vec_S}.
\end{eqnarray}
\end{subequations}
Note that $\vec{n}_A(t)$ is a vector operator acting on the bath only.
The comparison with Eq.\ (\ref{Htilde3}) yields
\begin{eqnarray}
 \label{nA_vs_psi_a}
\vec{n}_{A}(t) &=& 
\cos\psi(t) \vec{A}-\sin\psi(t)\left(\hat{a}(t)\times\vec{A}\right)
\nonumber
\\
&&+(1-\cos\psi(t))\hat{a}(t)(\hat{a}(t)\cdot\vec{A}).
\end{eqnarray}

\section{Expansion in $\tau_\mathrm{p} H$}
\label{Expansion}
We consider the case where the quantity $\tau_\mathrm{p} H$ is smaller than 
the pulse term $\tau_\mathrm{p} H_0$.  Then we expand in the small parameters
$\tau_\mathrm{p}\omega_\mathrm{b}$ and in $\tau_\mathrm{p}\lambda$, 
respectively, as they are defined after Eq.\ \eqref{Hamilt_bq}. 
Practically, we define the vector operator $\vec{A}(t)$ and 
we expand it in powers of $t$ \cite{pasin08a,pasin08b}
\begin{eqnarray}
 \label{A(Delta_tau)} \vec{A}(t)&:=&e^{\mathrm{i}H_\mathrm{b} t}\ \vec{A}\ 
e^{-\mathrm{i}H_\mathrm{b} t}
\nonumber\\
&=&\vec{A}+\sum_{n=1}^\infty\frac{(\mathrm{i}t)^n}{n!} 
[[H_\mathrm{b},\vec{A}]]_n.
\end{eqnarray} 
The notation $[[H_\mathrm{b},\vec{A}]]_n$ stands for the nested commutators 
$[H_\mathrm{b},[H_\mathrm{b},.....,[H_\mathrm{b},[H_\mathrm{b},\vec{A}]]]]$ 
with $H_\mathrm{b}$ appearing $n$ times.
Note that in the case of a static bath one has $[H_\mathrm{b},\vec{A}]=0$
and hence $\vec{A}(t) = \vec{A}$ holds.

According to Eqs.\
(\ref{Htilde},\ref{Hqb_vec_n},\ref{Hqb_vec_S}), the  time dependent
operator  $G(t)=\vec{S}(t)\cdot\vec{A}(t)=\vec{n}_{A(t)}(t)\cdot
\vec{\sigma}$  has the series
\begin{equation}
 \label{expansion_n(A)} 
G(t)=\vec{S}(t)\cdot\vec{A}+ \mathrm{i}t\ [H_\mathrm{b},
\vec{S}(t)\cdot\vec{A}]+O(t^2),
\end{equation} 
 in powers of $t$ or equivalently 
\begin{equation}
 \label{expansion_n(A)_1}
G(t)=\vec{n}_A(t)\cdot\vec{\sigma}+ \mathrm{i}t\ [H_\mathrm{b},
\vec{\sigma}\cdot\vec{n}_A(t)]+
O( t^2).
\end{equation}

Our aim is to make ${U}$, given formally in  \eqref{U_G}, 
as close as possible to the  identity operator.
We use the Magnus expansion \cite{magnu54} to eliminate the time ordering
in \eqref{U_G}. It reads 
\begin{equation}
\label{eq:magnus}
 {U}(\tau_\mathrm{p},0)=\exp\left[-\mathrm{i}
\tau_\mathrm{p}(G^{(1)}+G^{(2)}+...)
\right],
\end{equation}
with $\tau_\mathrm{p} G^{(1)}=\int_0^{\tau_\mathrm{p}} G(t) \mathrm{d}t$ and 
$\tau_\mathrm{p} G^{(2)}= -(\mathrm{i}/2)\int_0^{\tau_\mathrm{p}} 
\mathrm{d}t_1\int_0^{t_1}\mathrm{d}t_2[G(t_1),G(t_2)]$.
Combining the expansions \eqref{expansion_n(A)_1} and  \eqref{eq:magnus}
yields the wanted expansion in powers of $\tau_\mathrm{p}$ in the form
$U(\tau_\mathrm{p},0)=\exp\left[-\mathrm{i}(\eta^{(1)}+\eta^{(2)}+...)\right]$ 
where the first two terms are
\begin{equation}
 \label{eta1}
\eta^{(1)}=\int_0^{\tau_\mathrm{p}}\mathrm{d}t\ \vec{\sigma}\cdot\vec{n}_A(t)
\end{equation} 
\begin{eqnarray}
\label{eta2}
&&\eta^{(2)}=\mathrm{i}\int_0^{\tau_\mathrm{p}}\mathrm{d}t\ t\ [H_\mathrm{b}, 
\vec{\sigma}\cdot\vec{n}_A(t)]
\nonumber\\
&&-\frac{\mathrm{i}}{2}\int_0^{\tau_\mathrm{p}}\mathrm{d}t_1\int_0^{t_1}
\mathrm{d}t_2\ \left[\vec{n}_A(t_1)\cdot\vec{n}_A(t_2)-\vec{n}_A(t_2)
\cdot\vec{n}_A(t_1) \right.
\nonumber \\
&&\left.+\mathrm{i}\vec{\sigma}\cdot\left(\vec{n}_A(t_1)\times\vec{n}_A(t_2)
-\vec{n}_A(t_2)\times\vec{n}_A(t_1)\right)\right],
\end{eqnarray} 
with $\vec{n}_A(t)$ given by Eq.\ (\ref{nA_vs_psi_a}). 

We consider the most general 
case where $[\overline{A}_m,\overline{A}_n]\neq 0$ for $m\neq n$ with
$\vec{A}:=(\lambda_x \overline{A}_x,\lambda_y \overline{A}_y,\lambda_z 
\overline{A}_z)$. Equations \eqref{eta1} and \eqref{eta2}
are still operator equations. We wish to obtain a set of scalar
equations without operators. Hence we write 
$n_{A(t),i}(t)=\sum_{j=x,y,z} \lambda_jn_{i,j}(t) \overline{A}_j$, 
with $i=x,y,z$ and $\lambda_j$ being the strength of the coupling 
in the particular spin direction. The  
$n_{i,j}$ are operator independent scalars. They represent the matrix elements 
of the rotation matrix $D_{\hat{a}}(-\psi)$ as seen from the comparison of 
Eqs.\ (\ref{Hqb_vec_D}) and (\ref{Hqb_vec_S}). They are given
explicitly in the Appendix. These notations lead to the three expressions
\begin{equation}
\label{eq:matrix1}
\vec{\sigma}
\cdot\vec{n}_A(t)=\sum_{i,j}\sigma_i \lambda_jn_{i,j}(t)\overline{A}_j,
\end{equation}
\begin{eqnarray}
 \label{cross_product}
&&\left(\vec{n}_A(t_1)\times\vec{n}_A(t_2)-\vec{n}_A(t_2)
\times\vec{n}_A(t_1)\right)_i=
\nonumber\\
&&\sum_{j,k}\epsilon_{ijk}\sum_{l,m}\lambda_l\lambda_m
\left(\overline{A}_{l}\overline{A}_{m}+\overline{A}_{m}\overline{A}_{l}\right) 
n_{j,l}(t_1)n_{k,m}(t_2),
\nonumber\\
\end{eqnarray} 
\begin{eqnarray}
\label{scalar_product}
&&\vec{n}_A(t_1)\cdot\vec{n}_A(t_2)-\vec{n}_A(t_2)\cdot\vec{n}_A(t_1)=
\nonumber\\
&&\sum_{i;j<k}\lambda_j\lambda_k[\overline{A}_j,\overline{A}_k]
\left(n_{i,j}(t_1)n_{i,k}(t_2)-n_{i,j}(t_2)n_{i,k}(t_1)\right),
\nonumber\\
\end{eqnarray}
where  $\epsilon_{ijk}$ is the completely antisymmetric Levi-Civita tensor 
and $(\ )_i$ the component $i$ of the vector $(\ )$. 
Each index $i,\ j,\ k,\ l,\ m$ takes one of the values 
$x,\ y$, or $z$. Then 
Eqs.\ (\ref{eta1}) and (\ref{eq:matrix1}) imply 
\begin{equation}
 \label{eta1_components}
\eta^{(1)}=\sum_i\sigma_i \eta^{(1)}_i,
\end{equation}
with
\begin{eqnarray}
 \label{eta1_j} \eta^{(1)}_i:=
\sum_j\lambda_j \overline{A}_j\int_0^{\tau_\mathrm{p}}\mathrm{d}t\ n_{i,j}(t).
\end{eqnarray}
Eq.\ (\ref{eta2})  is conveniently split into
\begin{equation}
\eta^{(2)}=\sum_i\sigma_i\left(\eta^{(2a)}_i+
\eta^{(2b)}_{i}\right)+\eta^{(2c)}
\end{equation}
where
\begin{eqnarray}
 \label{eta2a_components} 
\eta^{(2a)}_i:=\sum_j\lambda_j[H_\mathrm{b},\overline{A}_j]
\int_0^{\tau_\mathrm{p}}\mathrm{d}t\ t\ n_{i,j}(t),
\end{eqnarray}
and with the help of \eqref{cross_product}
\begin{eqnarray}
 \label{eta2b_components} 
&&\eta^{(2b)}_{i}:=\sum_{l,m}\lambda_l\lambda_m\left(\overline{A}_{l}
\overline{A}_{m}+\overline{A}_{m}\overline{A}_{l}\right)\times
\nonumber\\
&&\int_0^{\tau_\mathrm{p}}\mathrm{d}t_1\int_0^{t_1}\mathrm{d}t_2\sum_{j,k}
\epsilon_{ijk} n_{j,l}(t_1)n_{k,m}(t_2)
\end{eqnarray} 
and with the help of \eqref{scalar_product}
\begin{eqnarray}
\label{eta2c_components}
&&\eta^{(2c)}:=\sum_{i;j<k}\lambda_j\lambda_k[\overline{A}_j,\overline{A}_k]
\times
\nonumber\\
&&\int_0^{\tau_\mathrm{p}}\mathrm{d}t_1\int_0^{t_1}\mathrm{d}t_2
\left(n_{i,j}(t_1)n_{i,k}(t_2)-n_{i,j}(t_2)n_{i,k}(t_1)\right).
\nonumber \\
\end{eqnarray}
If the components of $\vec{A}$ commute the correction $\eta^{(2c)}$ is zero
and  \eqref{eta2c_components} is fulfilled automatically.

\section{Discussion of the equations}
\label{Discussion} 
Approximating the ideal pulse, i.e., a $\delta$ peak, by the real pulse
$\vec{v}(t)$  up to the third order in $\Delta t$ is equivalent to imposing 
that the corrections $\eta^{(1)}$ and $\eta^{(2)}$ vanish. 
This implies the following system of integral equations 
\begin{subequations}
\label{gen_eqs}
\begin{equation}
\label{gen_eqs_eta1}
\int_0^{\tau_\mathrm{p}}\mathrm{d}t\ n_{i,j}(t)=0
\end{equation}
\begin{equation}
\label{gen_eqs_eta2a}
\int_0^{\tau_\mathrm{p}}\mathrm{d}t\ t\ n_{i,j}(t)=0
\end{equation}
for all $i,j\in\{x,y,z\}$ and
\begin{equation}
\label{gen_eqs_eta2b}
\iint_0^{\tau_\mathrm{p}}\mathrm{d}t_1\mathrm{d}t_2\sum_{j,k}
\epsilon_{ijk}n_{j,l}(t_1)n_{k,m}(t_2)\text{sgn}(t_1-t_2)=0.
\end{equation}
for all $i; l\le m$ and
\begin{equation}
\label{gen_eqs_eta2c}
\sum_i\iint_0^{\tau_\mathrm{p}}\mathrm{d}t_1\mathrm{d}t_2
n_{i,j}(t_1)n_{i,k}(t_2)\text{sgn}(t_1-t_2)=0,
\end{equation}
\end{subequations}
for $j<k$, i.e.,  for the
three cases $(j,k)=(x,y)$, $(j,k)=(x,z)$, and $(j,k)=(y,z)$. 
In the most general case, i.e., without
specific knowledge about the three operators $\overline{A}_{l}$ and
$H_\mathrm{b}$, the system consists of 39 integral equations.  But this 
number reduces drastically if specific cases are considered. 
Examples are a spin coupled to the bath only along the $z$ direction,
in which case Eq.\ (\ref{gen_eqs_eta2c}) can be neglected and 
\eqref{gen_eqs_eta2b} reduces to three equations, or 
if $[H_\mathrm{b},\vec{A}]=0$, in which case Eq.\ (\ref{gen_eqs_eta2a}) 
can be neglected. In the next sections we will 
provide and discuss solutions for such a specific case.

\renewcommand{\arraystretch}{1.2}
\begin{table}[t]
\begin{center}
\begin{tabular}{|c@{\extracolsep{0.1\columnwidth}}|c|}
\hline
amplitude(s) & $\tau_i$ \\
\hline
\multicolumn{2}{|c|}{\textbf{CORPSE-Pi}}\\
\hline
 $\pm 13\pi/6$      & 1/13 \\
                    & 6/13 \\
\hline
\multicolumn{2}{|c|}{\textbf{SCORPSE-Pi}}\\
\hline
 $\pm 7\pi/6$    & $1/7$ \\ 
                 & $6/7$ \\
\hline
\multicolumn{2}{|c|}{\textbf{SYM-Pi}}\\
\hline
 $\pm 17\pi /6$    & $5/17$ \\
                   & $12/17$ \\
\hline
\multicolumn{2}{|c|}{\textbf{SYM2ND-Pi}}\\
\hline
 $\pm 10.950120$  & 0.022805 \\
 $-7.695376$     & 0.275269 \\
                  & 0.724731 \\
                  & 0.977195 \\
\hline
\multicolumn{2}{|c|}{\textbf{ASYM2ND-Pi}}\\
\hline
 $\pm 11.364434$  & 0.252011 \\
                  & 0.310896 \\
                  & 0.584781 \\
                  & 0.752825 \\
                  & 0.796039 \\
\hline
\end{tabular}
\caption{Overview of the $\pi$ pulses satisfying all or parts of
the equations \eqref{gen_eqs}. SCORPSE-Pi,  
SYM-Pi and SYM2ND-Pi are symmetric pulses, while CORPSE-Pi and ASYM2ND-Pi are asymmetric. 
CORPSE-Pi, SCORPSE-Pi and SYM-Pi are pulses with vanishing first
order corrections ($\eta_{11}=\eta_{12}=0$), while SYM2ND-Pi and ASYM2ND-Pi make also the 
second order  corrections vanish ($\eta_{21}=\eta_{22}=\eta_{23}=0$). 
The switching instants $\tau_i$ and the amplitudes are given in units of 
$\tau_\mathrm{p}$ and $1/\tau_\mathrm{p}$, respectively. 
The SCORPSE-Pi coincides with UPi in Ref.\ \onlinecite{karba08}.
\label{tab_pulses_pi}}
\end{center}
\end{table}
\renewcommand{\arraystretch}{1}

\renewcommand{\arraystretch}{1.2}
\begin{table}[t]
\begin{center}
 \begin{tabular}{|c@{\extracolsep{0.1\columnwidth}}|c|}
\hline
amplitude(s) & $\tau_i$ \\
\hline
\multicolumn{2}{|c|}{\textbf{CORPSE-Pi2}}\\
\hline
  $\pm 6.345849$       & 0.033410  \\
                       & 0.471527   \\
\hline
\multicolumn{2}{|c|}{\textbf{SYM-Pi2}}\\
\hline
  $\pm 7.791318$       & 0.275201 \\
                       & 0.724799  \\
\hline
\multicolumn{2}{|c|}{\textbf{SYM2ND-Pi2}}\\
\hline
$\pm 11.486275$        & 0.037279  \\
$-8.038405$            & 0.269827  \\
                       & 0.730173  \\
                       & 0.962721  \\
\hline
\multicolumn{2}{|c|}{\textbf{ASYM2ND-Pi2}}\\
\hline
 $\pm 11.563810$  & 0.231411 \\
                  & 0.284623 \\
                  & 0.539588 \\
                  & 0.732138 \\
                  & 0.779722 \\
\hline
\end{tabular}
\caption{Overview of the $\pi/2$ pulses satisfying all or parts of
the equations \eqref{gen_eqs}. SYM-Pi2 and 
SYM2ND-Pi2 are symmetric pulses, while CORPSE-Pi2 and ASYM2ND-Pi2 are 
asymmetric. CORPSE-Pi2 and SYM-Pi2 are pulses with vanishing first
order corrections ($\eta_{11}=\eta_{12}=0$), while 
SYM2ND-Pi2 and ASYM2ND-Pi2 make also the second order corrections vanish 
($\eta_{21}=\eta_{22}=\eta_{23}=0$). The switching 
instants $\tau_i$ and the amplitudes are given in units of $\tau_\mathrm{p}$ 
and  $1/\tau_\mathrm{p}$, respectively. 
\label{tab_pulses_pi2}}
\end{center}
\end{table}
\renewcommand{\arraystretch}{1}

One of the advantages of the ansatz (\ref{ansatz0}) is that no free evolution 
occurs after the application of the pulse. Up to the corrected order in the 
expansion in $\tau_\mathrm{p}$, the effect of the pulse can be seen as being 
concentrated at the very end  of the  interval $[0,\tau_\mathrm{p}]$. 
This is surely a promising tool to be implemented 
experimentally. The possibility of designing pulses which effectively rotate 
the spin only at the very end of its duration  allows one to measure the 
response of a system without delay in time. This goal has been pursued
numerically in Refs.\ \onlinecite{gersh07} and \onlinecite{gersh08} for NMR 
pulses by optimal control theory. The spin dynamics was treated classically by 
the equation of motion for the magnetization; decoherence was included by a 
relaxation term.

In Refs.\  \onlinecite{pasin08a} and \onlinecite{pasin08b}, 
a no-go theorem for the second order correction of a $\pi$ pulse 
was proved for the ansatz \eqref{eq:ansatz-dd}. 
This limitation is eliminated for the ansatz \eqref{eq:present-goal}
considered in the present work. We present pulse shapes in the
following sections which make the first \emph{and all} 
the second order corrections vanish  as given in Eqs.\ (\ref{gen_eqs}), 
see Tables \ref{tab_pulses_pi} and \ref{tab_pulses_pi2}.

Pryadko and co-workers have previously studied a very similar issue 
\cite{sengu05,pryad08a,pryad08b}. They propose pulse shapes
which correct for the second order of static baths, i.e., baths
without internal dynamics \cite{sengu05}. Our pulses SYM2ND, ASYM2ND,
see Tables \ref{tab_pulses_pi} and \ref{tab_pulses_pi2}, 
and the continuous pulses, see Fig.\ \ref{fig_symm_cont_2nd}, go beyond
this level because they make also the second order corrections
vanish for dynamic baths. The corrections computed in Refs.\ 
\onlinecite{pryad08a} and \onlinecite{pryad08b} are special cases of our
equations \eqref{gen_eqs}: only symmetric pulses corresponding
to the rotation about a given spin axis are considered.

Some remarks about higher orders are in order. Third 
order corrections scale like $\omega_\mathrm{b}\lambda^2$, 
$\omega_\mathrm{b}^2\lambda$, or like $\lambda^3$. 
To make them vanish in the fully general framework, i.e.,
without prior knowledge of the operators $\vec{A}$, requires to fulfill
an additional finite number of equations.
Finding the corresponding solutions is a problem of a high degree of 
complexity because pulses with a more complicated structure must be considered.
Thus the numerical search of the corresponding
solutions becomes cumbersome. But there is no principal
reason not to tackle this issue in our approach considering
a bath with all quantum fluctuations, i.e., including all possible
effects of $J$ couplings in the NMR language. 

Before we proceed to the solutions, we mention
a geometrical interpretation of the corrections. 
Due to the identification of the $n_{i,j}$ with the matrix elements of the 
rotation matrix $D_{\hat{a}}(-\psi)$, one can 
interpret Eqs.\ (\ref{gen_eqs}) as the time average over all possible rotations
of the coupling term in the Hamiltonian (\ref{Hamilt_bq}) between the 
spin and the bath. This is particularly obvious in  Eq.\ (\ref{eta1_j}).
For the linear order analogous interpretations are quoted in the
literature for discrete sets of control pulses \cite{byrd02}.
The second order equations 
(\ref{gen_eqs_eta2a},\ref{gen_eqs_eta2b},\ref{gen_eqs_eta2c}) can be seen
as weighted averages of the $n_{i,j}$. An especially transparent geometric
interpretation has been reached if only the $z$ coupling $\overline{A}_z$
is present \cite{pasin08b}.

\section{Solutions}
\label{Solutions} 
In order to find solutions we focus on the simplest specific case.
We consider a spin coupled to the bath only along the 
$z$ direction, $\vec{A}=\lambda \overline{A}(0,0,1)$, and the pulse
consists only of a rotation around the  $y$ axis,  $\hat{a}=(0,1,0)$.
 Then the vector $\vec{n}_A$ is
\begin{equation}
 \label{nA_A_equal_z}\vec{n}:=\vec{n}^z=\lambda \overline{A}
\left(-\sin\psi(t),0,\cos\psi(t)\right).
\end{equation} 
Insertion in Eqs.\ (\ref{gen_eqs_eta1}, \ref{gen_eqs_eta2a}) 
and (\ref{gen_eqs_eta2b}) yields
\begin{subequations}
\label{etas_specific}
\begin{equation}
 \label{eta11_specialcase} \eta_{11}:=\int_0^{\tau_\mathrm{p}}
\mathrm{d}t \sin\psi(t)
\end{equation} 
\begin{equation}
 \label{eta12_specialcase} \eta_{12}:=\int_0^{\tau_\mathrm{p}}\mathrm{d}t 
\cos\psi(t)
\end{equation} 
\begin{equation}
 \label{eta21_specialcase} \eta_{21}:=\int_0^{\tau_\mathrm{p}}\mathrm{d}t\ t\ 
\sin\psi(t)
\end{equation} 
\begin{equation}
 \label{eta22_specialcase} \eta_{22}:=\int_0^{\tau_\mathrm{p}}\mathrm{d}t\ t\ 
\cos\psi(t)
\end{equation} 
\begin{equation}
 \label{eta23_specialcase} \eta_{23}:=\iint_0^{\tau_\mathrm{p}}
\mathrm{d}t_1\mathrm{d}t_2 \sin(\psi({t_1})-\psi({t_2}))\text{sgn}(t_1-t_2).
\end{equation} 
\end{subequations}
For the first order corrections to vanish 
$\eta_{11}=\eta_{12}=0$ must hold. For
the second order corrections also to vanish
$\eta_{21}=\eta_{22}=\eta_{23}=0$ is required in addition.

\begin{figure}
\begin{center}
     \includegraphics[width=\columnwidth,clip]{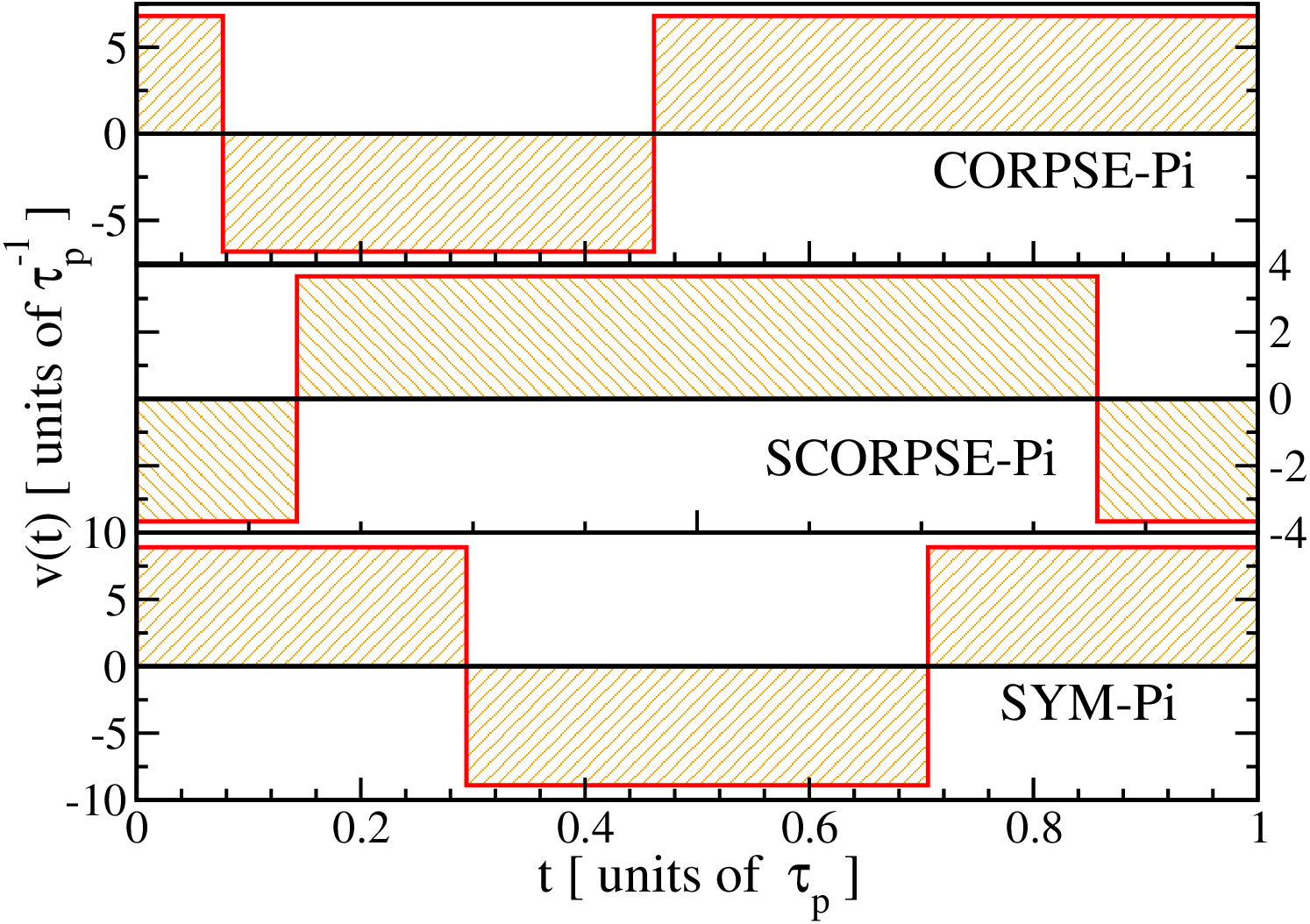}
\end{center}
\caption{(Color online)
Symmetric and asymmetric $\pi$ pulses with piecewise constant 
amplitude. The pulse characteristics are reported in Table \ref{tab_pulses_pi}.
 These pulses make the first order corrections 
Eqs.\ (\ref{eta11_specialcase}) and (\ref{eta12_specialcase}) vanish. 
The pulses CORPSE  and SCORPSE coincide with those proposed in 
\cite{cummi00,cummi03}.  The 
SCORPSE-Pi coincides with UPi in Ref.\ \onlinecite{karba08}.}
\label{fig_Pi_1ord}
\end{figure}
\begin{figure}
\begin{center}
     \includegraphics[width=\columnwidth,clip]{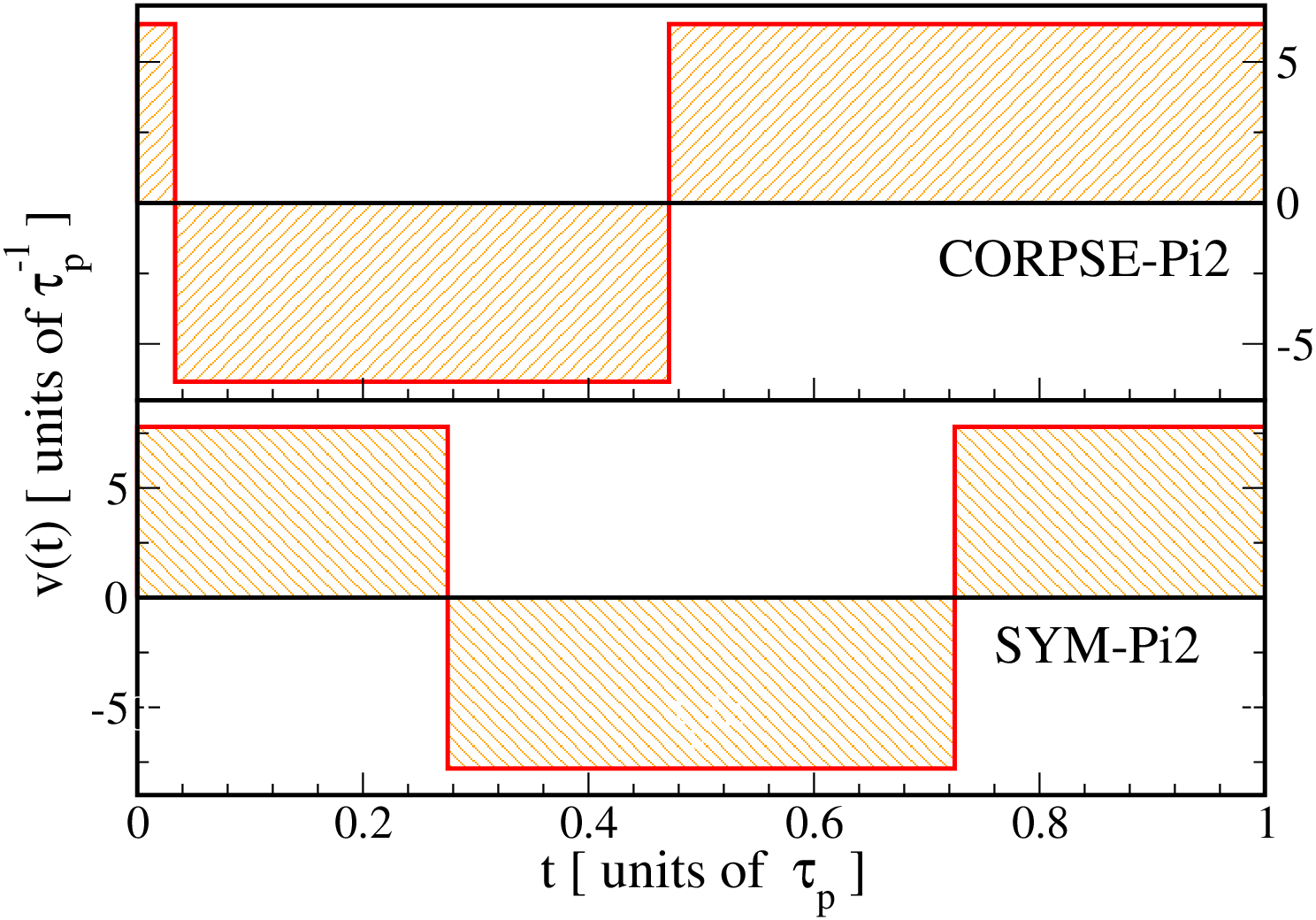}
\end{center}
\caption{(Color online)
Symmetric and asymmetric $\pi/2$ pulses with piecewise constant 
amplitude. The pulse characteristics are reported in Table \ref{tab_pulses_pi}.
 These pulses make the first order corrections Eq.\ (\ref{eta11_specialcase}) 
and (\ref{eta12_specialcase}) vanish. The pulse CORPSE coincides with that 
proposed in Refs.\ \onlinecite{cummi00} and \onlinecite{cummi03}.}
\label{fig_Pi2_1ord}
\end{figure}

\subsection{Composite pulses}
A composite pulse is a pulse which can be seen as being composed
of simple pulses of constant amplitudes, i.e., the total pulse
is characterized by piecewise constant amplitudes.
\footnote{Sometimes the term `composite' pulse is used for any concatenation
of pulses. We prefer to stick to the more restricted use of
composite pulses for pulses of piecewise constant amplitude because
in a less strict notation it is not possible to distinguish rigorously
between a simple and a composite pulse.}
For simplicity, we parametrize the pulse shape as function of constant modulus 
of the amplitude and search for
solutions to Eqs.\ (\ref{etas_specific}) for $\pi$ and $\pi/2$ pulses. 

\begin{figure}
\begin{center}
     \includegraphics[width=\columnwidth,clip]{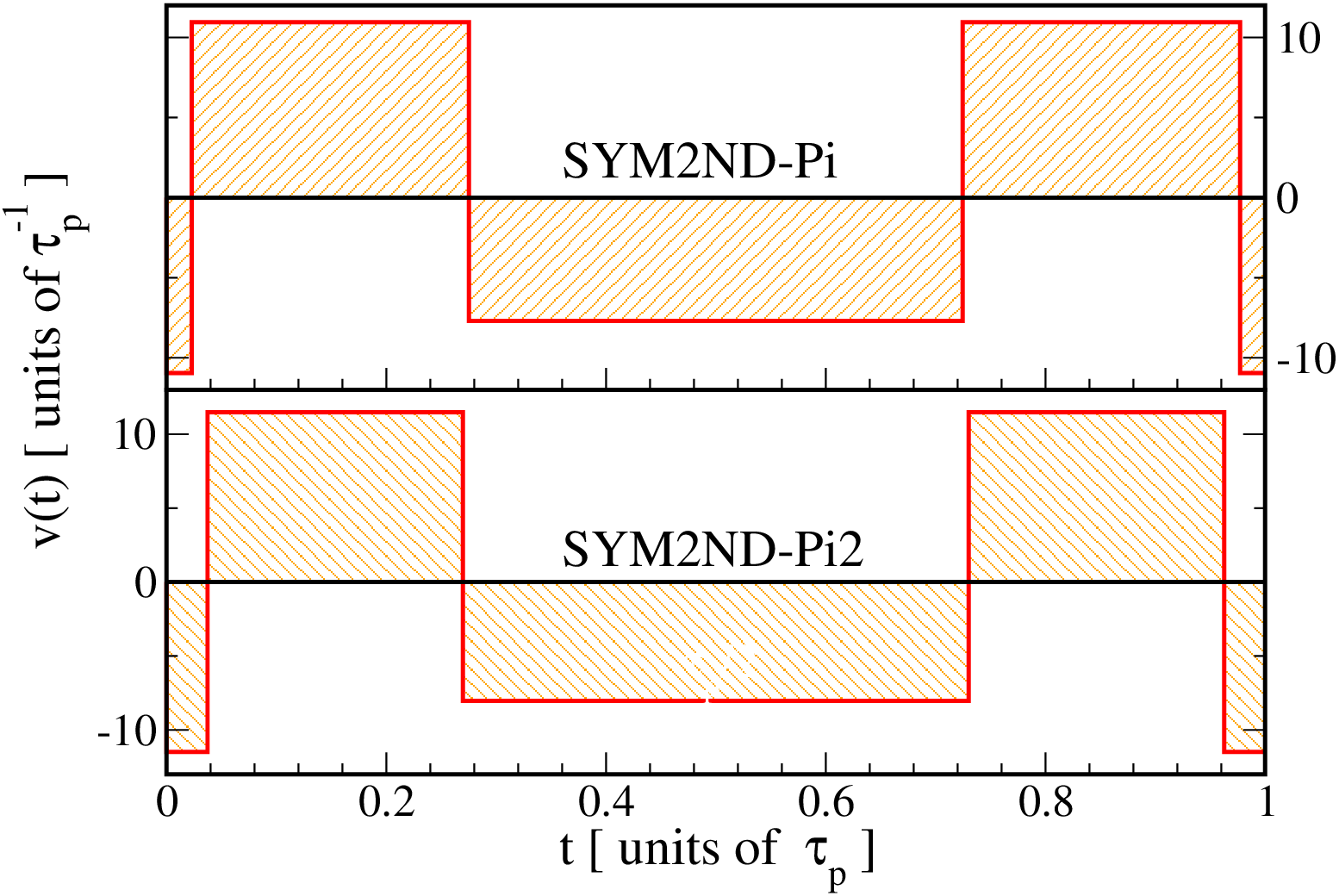}
\end{center}
\caption{(Color online)
Symmetric $\pi$ and $\pi/2$ pulse with piecewise constant 
amplitude that correct also the second order 
($\eta_{21}=\eta_{22}=\eta_{23}=0$), see
Eqs.\ (\ref{eta21_specialcase},\ref{eta22_specialcase}) and 
Eq.\ \eqref{eta23_specialcase}. Their characteristics are reported in Table 
\ref{tab_pulses_pi} and in Table \ref{tab_pulses_pi2}.}
\label{fig_2ord_sym}
\end{figure}

\begin{figure}
\begin{center}
     \includegraphics[width=\columnwidth,clip]{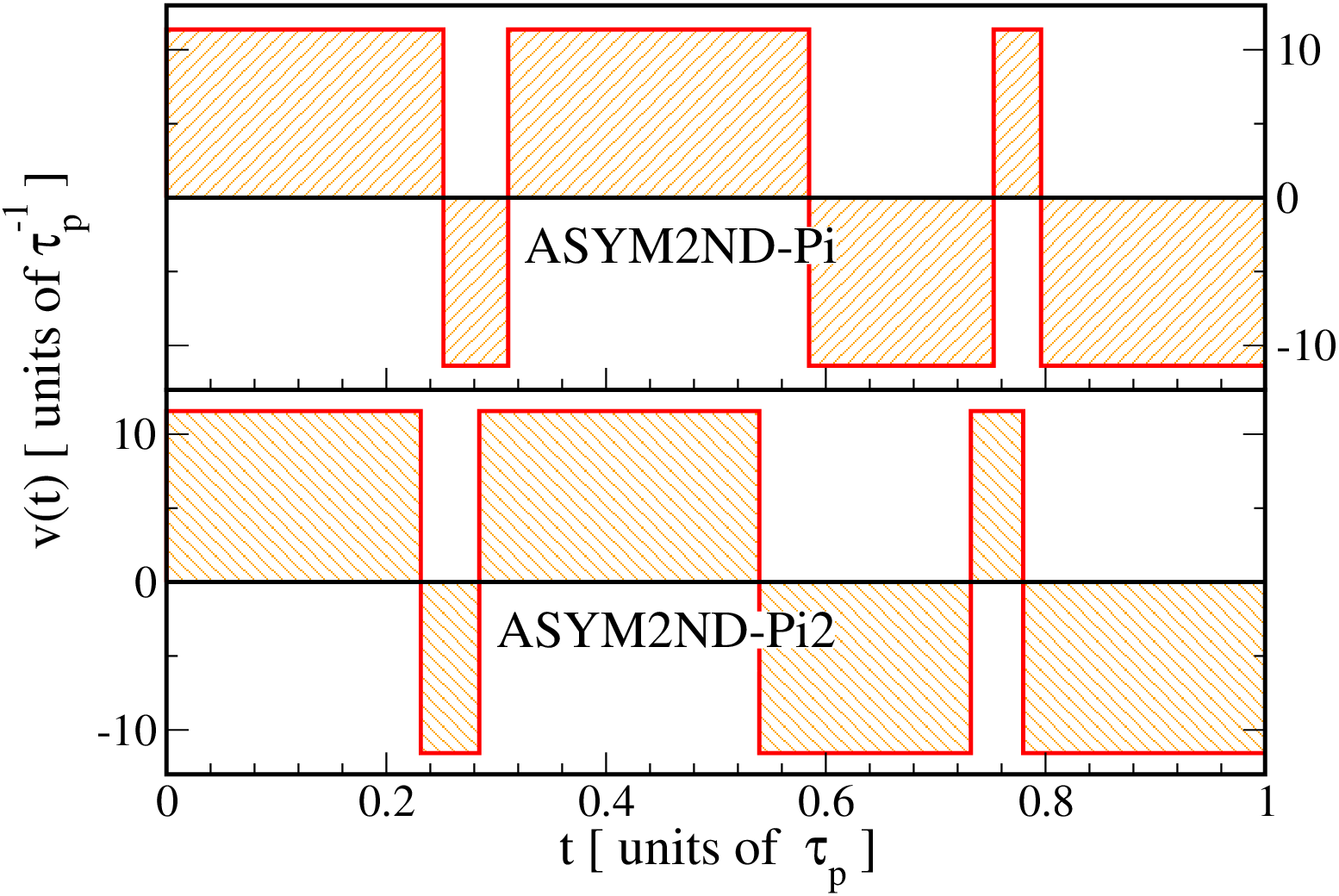}
\end{center}
\caption{(Color online)
Asymmetric $\pi$ and $\pi/2$ pulse with piecewise constant 
amplitude that correct also the second order 
($\eta_{21}=\eta_{22}=\eta_{23}=0$), see
Eqs.\ (\ref{eta21_specialcase},\ref{eta22_specialcase}) and 
Eq.\ \eqref{eta23_specialcase}. Their characteristics are reported in Table 
\ref{tab_pulses_pi} and in Table \ref{tab_pulses_pi2}.}
\label{fig_2ord_asym}
\end{figure}

If  we restrict ourselves to the first order corrections 
we find that the CORPSE pulses satisfy our equations, 
both for the $\pi$ and $\pi/2$ case. The definition of the
SCORPSE pulse \cite{cummi03} works only in the $\pi$ case
(with flipped amplitudes, see Ref.\ \onlinecite{motto06}). 
For other angles $\theta$ it actually yields the angle $\theta-2\pi$.
Amplitudes and switching instants 
are reported in Tables \ref{tab_pulses_pi} and  \ref{tab_pulses_pi2}. 
The first order correcting pulses are plotted in Figs.\ \ref{fig_Pi_1ord} 
and \ref{fig_Pi2_1ord} while the pulses correcting also the second order are 
displayed in 
Figs.\ \ref{fig_2ord_sym} and \ref{fig_2ord_asym}. 
The characteristics of the pulses are given in 
Tables \ref{tab_pulses_pi} and \ref{tab_pulses_pi2}.
Note that these solutions are
qualitatively very similar to the very short pulses found numerically by
optimum control theory applied to the classical dynamics of the
magnetization in NMR \cite{kobza04}.

CORPSE pulses are asymmetric pulses while SCORPSE pulses are symmetric. 
These results confirm that our ansatz for the decoupling of the spin from 
the bath comprises the ansatz that led to the  CORPSE and the SCORPSE pulse. 
In this sense our results reproduce those in Refs.\  
\onlinecite{cummi00,cummi03,motto06} generally and they reach beyond
the former references  because (i) the bath is treated
quantum mechanically and (ii) a larger variety of pulses is considered, see 
for example SYM-Pi and SYM-Pi2 in 
Tables \ref{tab_pulses_pi} and \ref{tab_pulses_pi2}.

Alway and Jones \cite{alway07} derived composite pulses 
along the lines proposed by Brown {\it et al.} \cite{brown04}.
As far as off-resonance errors are considered the wanted
pulse properties are the same at which we are aiming.
But the correction in second order works only for purely
static baths, i.e., for $H_\mathrm{b}=0$. Note that fairly complicated
pulses are required with fine-tuned axes of rotation while
our second order pulses SYM2ND-Pi and SYM2ND-Pi2
do with a fixed axis of rotation. Furthermore, the approach used in Refs.\ 
\onlinecite{alway07} and \onlinecite{brown04}
is suited only for composite pulses.


\subsection{Continuous pulses}

In a setup where frequency selectivity is required,
for instance for magnetic resonance imaging (MRI) \cite{ernst87}
the piecewise constant pulses, i.e., the composite pulses,
are not the optimum choice due to their jumps. For this reason, we also 
propose continuous pulses which are characterized by narrower frequency bands. 
The pulses we obtain are qualitatively similar in their smoothness
to the pulses of intermediate duration found numerically by
optimum control theory applied to the classical dynamics of the
magnetization in NMR \cite{kobza04}.

\begin{figure}
\begin{center}
     \includegraphics[width=\columnwidth,clip]{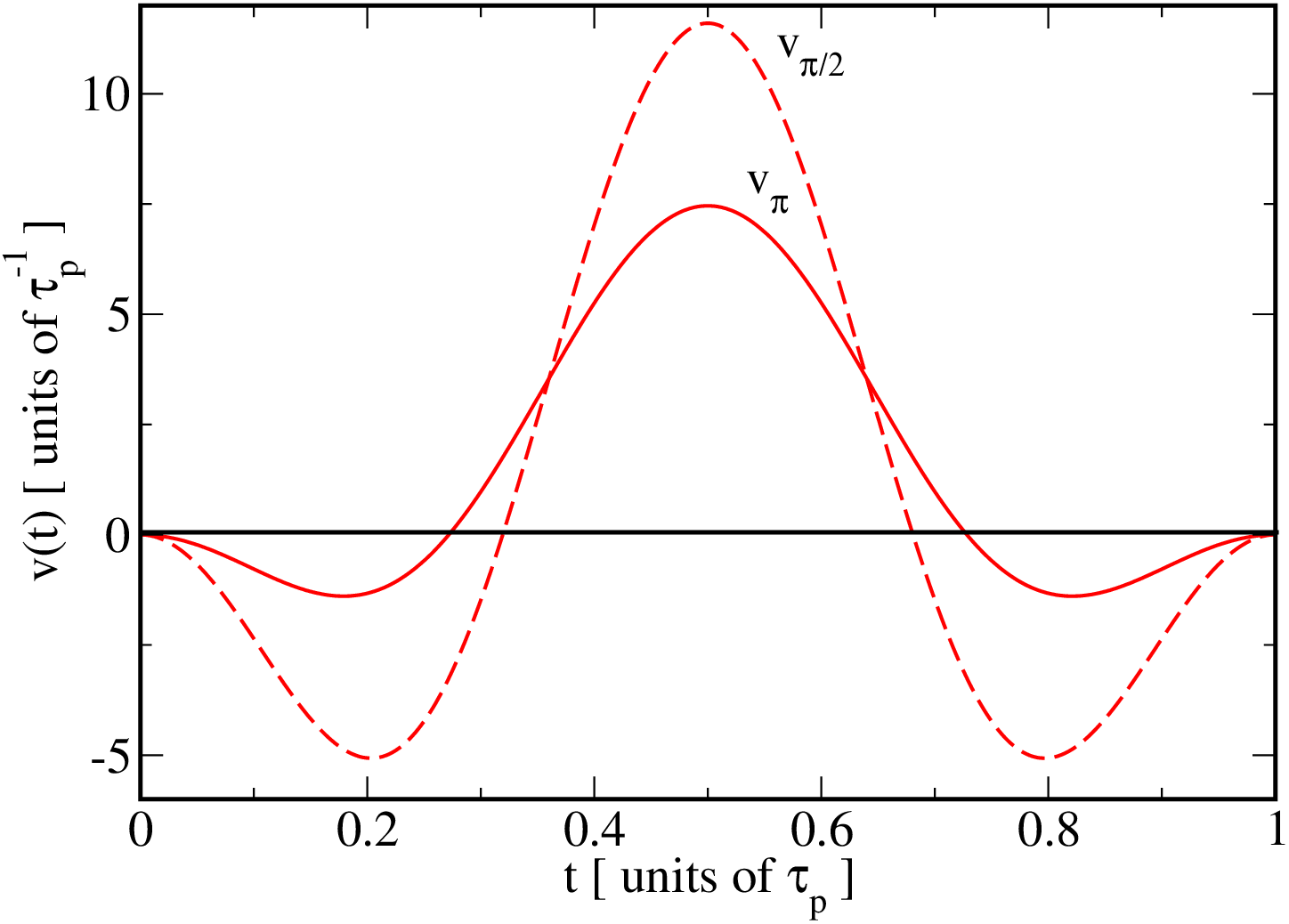}
\end{center}
\caption{(Color online)
Symmetric $\pi$ and $\pi/2$ pulses with continuous amplitude. 
Both are solutions of the first order Eqs.\ (\ref{eta11_specialcase}) and 
(\ref{eta12_specialcase}). We refer to Eq.\ (\ref{ansatz-cont-sym}) for their 
parametrization. For the $\pi$ pulse we find $a=-2.159224[1/\tau_\mathrm{p}]$, 
which coincides with the pulse proposed in Ref.\ \onlinecite{pasin08a}
as expected.
For the $\pi/2$ pulse we  find $a=-5.015588[1/\tau_\mathrm{p}]$ which differs
from previous proposals since another approximation is pursued, 
see ansatz \eqref{eq:present-goal}. The maximum amplitude is given by 
$a_\mathrm{max}=(\theta -2 a)[1/\tau_\mathrm{p}]$.}
\label{fig_symm_cont}
\end{figure}

\begin{figure}
\begin{center}
     \includegraphics[width=\columnwidth,clip]{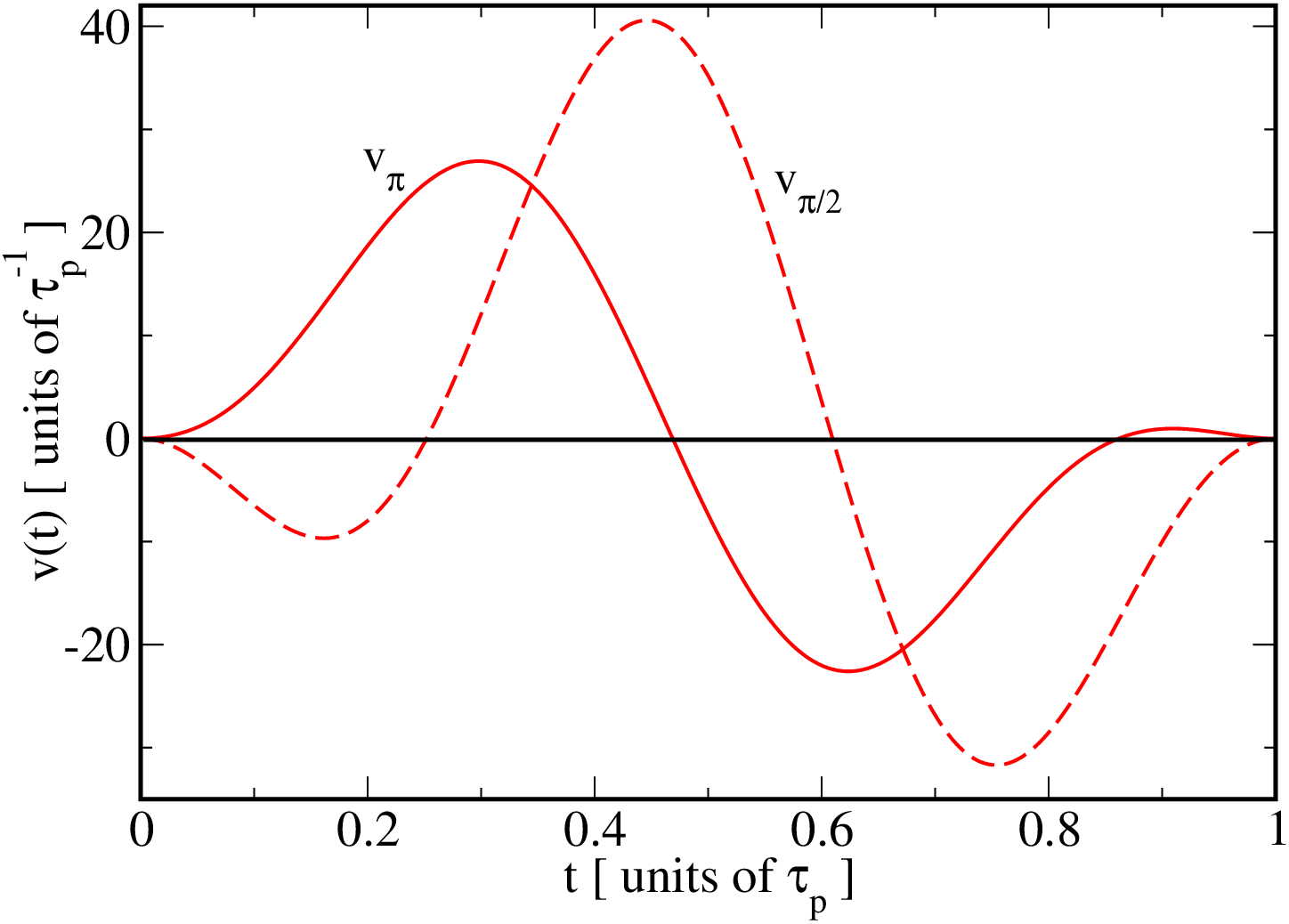}
\end{center}
\caption{(Color online)
Asymmetric $\pi$ and $\pi/2$ pulses with continuous amplitude. 
Both are solutions of the first order  Eqs.\ (\ref{eta11_specialcase}) and 
(\ref{eta12_specialcase}). We refer to Eq.\ (\ref{ansatz-cont-asym}) for their 
parametrization. For the $\pi$ pulse we find $a=5.263022[1/\tau_\mathrm{p}]$ 
and  $b=17.850535[1/\tau_\mathrm{p}]$ while for the $\pi/2$ pulse  we find 
$a=-16.809353[1/\tau_\mathrm{p}]$ and $b=15.634390[1/\tau_\mathrm{p}]$. The 
maximum amplitudes are $a_\mathrm{max}=26.916283[1/\tau_\mathrm{p}]$ at 
$t=0.2977 \tau_\mathrm{p}$ for the $\pi$ pulse and 
$a_\mathrm{max}=40.572755[1/\tau_\mathrm{p}]$ at $t=0.4461\tau_\mathrm{p}$ for 
the $\pi/2$ pulse.}
\label{fig_asymm_cont}
\end{figure}

First, we study the first order corrections. As before we design $\pi$ 
and $\pi/2$ pulses without the linear corrections: $\eta_{11}=\eta_{12}=0$.
For  symmetric pulses the function $v(t)$ can be expanded in the 
Fourier series  
\begin{equation}
\label{ansatz-cont-sym}
 v(t)={\theta}/{2}+(a-\theta/2)\cos(2\pi t/\tau_\mathrm{p})-a
\cos(4\pi t/\tau_\mathrm{p}),
\end{equation}
with $\theta$ being either equal to $\pi$ or to $\pi/2$.
The amplitude $a$ is the parameter which is varied to comply with
 $\eta_{11}=\eta_{12}=0$. For asymmetric pulses we choose
\begin{eqnarray}
\nonumber
v(t) &=& {\theta}/{2}+(a-\theta/2)\cos(2\pi t/\tau_\mathrm{p})-a 
\cos(4\pi t/\tau_\mathrm{p})
\\
&& +b\sin(2\pi t/\tau_\mathrm{p}) - ({b}/{2})\sin(4\pi t/\tau_\mathrm{p}),
\label{ansatz-cont-asym}
\end{eqnarray} 
where $a$ and $b$ are varied to reach $\eta_{11}=\eta_{12}=0$.
The symmetric pulses are plotted in Fig.\ \ref{fig_symm_cont} and 
the asymmetric ones in Fig.\ \ref{fig_asymm_cont}.

For closed systems consisting of several Ising spins in a piece of chain,
but without any  bath $H_\mathrm{b}$, Sengupta and Pryadko \cite{sengu05} 
proposed tuned symmetric pulses making the first and the 
second order corrections vanish. The pulses are designed such that 
the  $2L-1$ first derivatives of the pulse amplitude  with $L\in\{1,2\}$ are 
zero at $t=0$ and $t=\tau_\mathrm{p}$. We checked that all these pulses make 
the first order corrections in Eqs.\ 
(\ref{eta11_specialcase},\ref{eta12_specialcase}) vanish.

Concerning the second order corrections, the pulses suggested in Ref.\
\onlinecite{sengu05} make the term $\eta_{23}$ vanish, see 
Eq.\ (\ref{eta23_specialcase}), but not the 
terms $\eta_{21}$ and $\eta_{22}$ in 
Eqs.\ (\ref{eta21_specialcase},\ref{eta22_specialcase}). 
This is consistent with the fact that terms proportional
to $\eta_{21}$ and $\eta_{22}$ do not appear in the second order 
corrections if no explicit bath dynamics $H_\mathrm{b}$ is present, see 
Eq.\ (\ref{eta2a_components}).


\begin{figure}
\begin{center}
     \includegraphics[width=\columnwidth,clip]{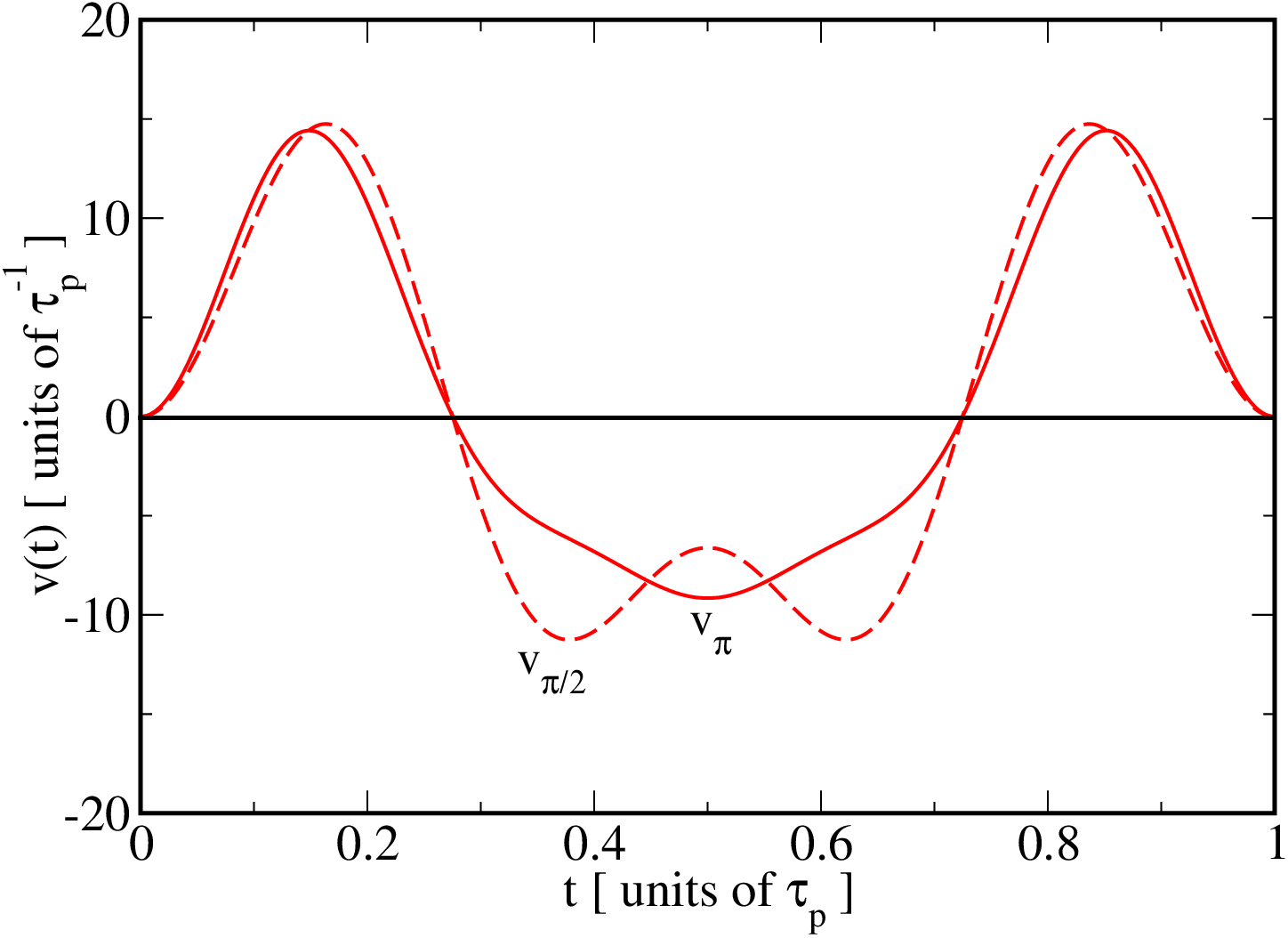}
\end{center}
\caption{(Color online)
Symmetric $\pi$ and $\pi/2$ pulses with continuous amplitudes. 
Both are solutions of the first and second order  
Eqs.\ (\ref{etas_specific}). 
We refer to Eq.\ (\ref{eq:continuous}) for their 
parametrization. For the $\pi$ pulse we find $a=10.804433[1/\tau_\mathrm{p}]$,
$b=6.831344[1/\tau_\mathrm{p}]$, and $c=2.174538[1/\tau_\mathrm{p}]$
 while for the $\pi/2$ pulse we find $a=10.925826[1/\tau_\mathrm{p}]$,
$b=6.806775[1/\tau_\mathrm{p}]$, and $c=-0.02696178[1/\tau_\mathrm{p}]$.}
\label{fig_symm_cont_2nd}
\end{figure}

We succeeded also to find continuous pulses which make the
first and the second order terms vanish, i.e., which fulfill
the whole set of equations \eqref{etas_specific}. We use
the symmetric ansatz
\begin{eqnarray}
 v(t)\!\!&=&\!\!
{\theta}/{2}+(a-\theta/2)\cos(2\pi t/\tau_\mathrm{p})+(b-a)
\cos(4\pi t/\tau_\mathrm{p})\nonumber\\
&& +(c-b) \cos(6\pi t/\tau_\mathrm{p}) - c \cos(8\pi t/\tau_\mathrm{p}),
\label{eq:continuous}
\end{eqnarray}
where $a$, $b$, and $c$ are the amplitudes to be determined.
The resulting pulses are displayed in Fig.\ \ref{fig_symm_cont_2nd}.
Interestingly, the pulses are fairly simple in structure with only
two zeros between 0 and 1. 

An additional explanation on the band width of these pulses is
in order. Of course, the fact that our pulses
make the first and the second order vanish implies
that they constitute robust broadband pulses. At first
sight, this contradicts the use as frequency selective pulses.
But in our simulations (not shown here) we find that the
applicability of the second order correcting pulses
vanishes very quickly as function of the detuning. Hence
they display a very good selectivity for long enough pulses, i.e,,
large enough $\tau_\text{p}$.
The fact that they even partly compensate the dynamics
of the decoherence bath makes them robust. Hence they are
promising candidates for the application in MRI.

\section{Conclusions}
\label{Conclusions} 

In this article we aimed at a complete decoupling of a spin from 
its quantum mechanical bath, induced for instance by other nuclear spins
through hyperfine couplings,
 during a general short control pulse. This aim was different
from the one we followed previously \cite{pasin08a,karba08,pasin08b}
where we aimed at the disentanglement of the Hamiltonians of the
pulse and of the system. The aim followed in the present work was closer
to the aims pursued in the literature.

We studied a general pulse applied to a general spin-bath model in
a perturbation approach in the shortness of the pulse duration 
$\tau_\mathrm{p}$. The first and second order correction terms have been 
derived generally. This set of equations includes many results of previous 
investigations such as, for instance, off-resonance models where the 
perturbing part is static. Pulses like the CORPSE pulse 
\cite{cummi00,cummi03,motto06} are shown to make the first
order corrections vanish but not the general second order corrections.

In the specific example of a model with dephasing bath, which shows 
internal dynamics,
we demonstrated the existence of pulses which make the first and the 
second order corrections vanish. To our knowledge, no such result has
been presented for quantum mechanical baths
so far in the literature. This finding illustrates that the
no-go theorem concerning the disentanglement of $\pi$ pulses
from the system \cite{pasin08a,pasin08b} does not apply if one aims
at the perturbative decoupling of spin and system.

The present results are useful for applications in quantum information
processing (QIP), nuclear magnetic resonance (NMR), and 
magnetic resonance imaging (MRI). In the framework of QIP,
single quantum bit gates can be realized reliably by control pulses which
fulfill the equations derived here. In the framework of NMR, the 
measurement of the time evolution of certain quantum states can
be performed after their preparation without time delay because the real
pulse behaves like an instantaneous pulse at the very end of
its finite duration. We emphasize that our result
extends previous ones \cite{skinn03,kobza04,luy05,gersh07,gersh08} 
in the sense that a decoherence bath with quantum mechanical dynamics is
considered ($J$ couplings in NMR).
In the framework of MRI, the continuous pulses correcting the first and
the second order are potential candidates for frequency selective
pulses.
Last but not least, the pulses proposed here can be applied
for the realization of especially adapted sequences for dynamic decoupling
\cite{uhrig09c}.

\begin{acknowledgments}
We thank G.~Alvarez, J.~A.\ Jones, L.~P.\ Pryadko, T.~E. Skinner,
and D.~Suter for many useful comments.
\end{acknowledgments}

\bigskip

\section{Appendix}
\label{appendix}
We start from $\vec{n}_A(t)=D_{\hat{a}}(-\psi)\vec{A}$
where $D_{\hat{a}}(-\psi)$ stands for the matrix representing 
the rotation around the unit vector $\hat{a}(t)$ by the angle $\psi(t)$. 
From the explicit form of $\vec{n}_A(t)$, Eq.\ (\ref{nA_vs_psi_a}), 
one can derive the matrix elements of 
$D_{\hat{a}}(-\psi)$ which we introduced as  $n_{i,j}$
\begin{widetext}
 \begin{equation}D_{\hat{a}}(-\psi)=\left(
 \begin{array}{ccc}  
\cos\psi+(1-\cos\psi)a_x^2 &  a_z\sin\psi+(1-\cos\psi)a_xa_y & -a_y\sin\psi+
(1-\cos\psi)a_xa_z \\ 
-a_z\sin\psi+(1-\cos\psi)a_xa_y &  \cos\psi+(1-\cos\psi) a_y^2 & a_x\sin\psi+
(1-\cos\psi)a_ya_z \\
a_y\sin\psi+(1-\cos\psi)a_xa_z & -a_x\sin\psi+(1-\cos\psi)a_ya_z & \cos\psi+
(1-\cos\psi)a_z^2
 \end{array}\right)
\end{equation}\end{widetext} where the time dependence of $\psi(t)$ and 
$\hat{a}(t)$ has been omitted for clarity.


\begin{thebibliography}{37}
\expandafter\ifx\csname natexlab\endcsname\relax\def\natexlab#1{#1}\fi
\expandafter\ifx\csname bibnamefont\endcsname\relax
  \def\bibnamefont#1{#1}\fi
\expandafter\ifx\csname bibfnamefont\endcsname\relax
  \def\bibfnamefont#1{#1}\fi
\expandafter\ifx\csname citenamefont\endcsname\relax
  \def\citenamefont#1{#1}\fi
\expandafter\ifx\csname url\endcsname\relax
  \def\url#1{\texttt{#1}}\fi
\expandafter\ifx\csname urlprefix\endcsname\relax\def\urlprefix{URL }\fi
\providecommand{\bibinfo}[2]{#2}
\providecommand{\eprint}[2][]{\url{#2}}

\bibitem[{\citenamefont{Hahn}(1950)}]{hahn50}
\bibinfo{author}{\bibfnamefont{E.~L.} \bibnamefont{Hahn}},
  \bibinfo{journal}{Phys. Rev.} \textbf{\bibinfo{volume}{80}},
  \bibinfo{pages}{580} (\bibinfo{year}{1950}).

\bibitem[{\citenamefont{Carr and Purcell}(1954)}]{carr54}
\bibinfo{author}{\bibfnamefont{H.~Y.} \bibnamefont{Carr}} \bibnamefont{and}
  \bibinfo{author}{\bibfnamefont{E.~M.} \bibnamefont{Purcell}},
  \bibinfo{journal}{Phys. Rev.} \textbf{\bibinfo{volume}{94}},
  \bibinfo{pages}{630} (\bibinfo{year}{1954}).

\bibitem[{\citenamefont{Meiboom and Gill}(1958)}]{meibo58}
\bibinfo{author}{\bibfnamefont{S.}~\bibnamefont{Meiboom}} \bibnamefont{and}
  \bibinfo{author}{\bibfnamefont{D.}~\bibnamefont{Gill}},
  \bibinfo{journal}{Rev. Sci. Inst.} \textbf{\bibinfo{volume}{29}},
  \bibinfo{pages}{688} (\bibinfo{year}{1958}).

\bibitem[{\citenamefont{Haeberlen}(1976)}]{haebe76}
\bibinfo{author}{\bibfnamefont{U.}~\bibnamefont{Haeberlen}},
  \emph{\bibinfo{title}{High Resolution NMR in Solids: Selective Averaging}}
  (\bibinfo{publisher}{Academic Press}, \bibinfo{address}{New York},
  \bibinfo{year}{1976}).

\bibitem[{\citenamefont{Ernst et~al.}(1987)\citenamefont{Ernst, Bodenhausen,
  and Wokaun}}]{ernst87}
\bibinfo{author}{\bibfnamefont{R.~R.} \bibnamefont{Ernst}},
  \bibinfo{author}{\bibfnamefont{G.}~\bibnamefont{Bodenhausen}},
  \bibnamefont{and} \bibinfo{author}{\bibfnamefont{A.}~\bibnamefont{Wokaun}},
  \emph{\bibinfo{title}{Principles of Nuclear Magnetic Resonance in One and Two
  Dimensions}}, vol.~\bibinfo{volume}{14} of
  \emph{\bibinfo{series}{International Series of Monographs on Chemistry}}
  (\bibinfo{publisher}{Clarendon Press}, \bibinfo{address}{Oxford},
  \bibinfo{year}{1987}).

\bibitem[{\citenamefont{Freeman}(1998)}]{freem98}
\bibinfo{author}{\bibfnamefont{R.}~\bibnamefont{Freeman}},
  \emph{\bibinfo{title}{Spin Choreography: Basic Steps in High Resolution NMR}}
  (\bibinfo{publisher}{Oxford University Press}, \bibinfo{address}{Oxford},
  \bibinfo{year}{1998}).

\bibitem[{\citenamefont{Viola and Lloyd}(1998)}]{viola98}
\bibinfo{author}{\bibfnamefont{L.}~\bibnamefont{Viola}} \bibnamefont{and}
  \bibinfo{author}{\bibfnamefont{S.}~\bibnamefont{Lloyd}},
  \bibinfo{journal}{Phys. Rev. A} \textbf{\bibinfo{volume}{58}},
  \bibinfo{pages}{2733} (\bibinfo{year}{1998}).

\bibitem[{\citenamefont{Ban}(1998)}]{ban98}
\bibinfo{author}{\bibfnamefont{M.}~\bibnamefont{Ban}}, \bibinfo{journal}{J.
  Mod. Opt.} \textbf{\bibinfo{volume}{45}}, \bibinfo{pages}{2315}
  (\bibinfo{year}{1998}).

\bibitem[{\citenamefont{Viola et~al.}(1999)\citenamefont{Viola, Knill, and
  Lloyd}}]{viola99a}
\bibinfo{author}{\bibfnamefont{L.}~\bibnamefont{Viola}},
  \bibinfo{author}{\bibfnamefont{E.}~\bibnamefont{Knill}}, \bibnamefont{and}
  \bibinfo{author}{\bibfnamefont{S.}~\bibnamefont{Lloyd}},
  \bibinfo{journal}{Phys. Rev. Lett.} \textbf{\bibinfo{volume}{82}},
  \bibinfo{pages}{2417} (\bibinfo{year}{1999}).

\bibitem[{\citenamefont{Khodjasteh and Lidar}(2005)}]{khodj05}
\bibinfo{author}{\bibfnamefont{K.}~\bibnamefont{Khodjasteh}} \bibnamefont{and}
  \bibinfo{author}{\bibfnamefont{D.~A.} \bibnamefont{Lidar}},
  \bibinfo{journal}{Phys. Rev. Lett.} \textbf{\bibinfo{volume}{95}},
  \bibinfo{pages}{180501} (\bibinfo{year}{2005}).

\bibitem[{\citenamefont{Uhrig}(2007)}]{uhrig07}
\bibinfo{author}{\bibfnamefont{G.~S.} \bibnamefont{Uhrig}},
  \bibinfo{journal}{Phys. Rev. Lett.} \textbf{\bibinfo{volume}{98}},
  \bibinfo{pages}{100504} (\bibinfo{year}{2007}).

\bibitem[{\citenamefont{Lee et~al.}(2008)\citenamefont{Lee, Witzel, and
  \mbox{Das~Sarma}}}]{lee08a}
\bibinfo{author}{\bibfnamefont{B.}~\bibnamefont{Lee}},
  \bibinfo{author}{\bibfnamefont{W.~M.} \bibnamefont{Witzel}},
  \bibnamefont{and}
  \bibinfo{author}{\bibfnamefont{S.}~\bibnamefont{Das~Sarma}},
  \bibinfo{journal}{Phys. Rev. Lett.} \textbf{\bibinfo{volume}{100}},
  \bibinfo{pages}{160505} (\bibinfo{year}{2008}).

\bibitem[{\citenamefont{Yang and Liu}(2008)}]{yang08}
\bibinfo{author}{\bibfnamefont{W.}~\bibnamefont{Yang}} \bibnamefont{and}
  \bibinfo{author}{\bibfnamefont{R.-B.} \bibnamefont{Liu}},
  \bibinfo{journal}{Phys. Rev. Lett.} \textbf{\bibinfo{volume}{101}},
  \bibinfo{pages}{180403} (\bibinfo{year}{2008}).

\bibitem[{\citenamefont{Biercuk
  et~al.}(2009{\natexlab{a}})\citenamefont{Biercuk, Uys, VanDevender, Shiga,
  Itano, and Bollinger}}]{bierc09a}
\bibinfo{author}{\bibfnamefont{M.~J.} \bibnamefont{Biercuk}},
  \bibinfo{author}{\bibfnamefont{H.}~\bibnamefont{Uys}},
  \bibinfo{author}{\bibfnamefont{A.~P.} \bibnamefont{VanDevender}},
  \bibinfo{author}{\bibfnamefont{N.}~\bibnamefont{Shiga}},
  \bibinfo{author}{\bibfnamefont{W.~M.} \bibnamefont{Itano}}, \bibnamefont{and}
  \bibinfo{author}{\bibfnamefont{J.~J.} \bibnamefont{Bollinger}},
  \bibinfo{journal}{Nature} \textbf{\bibinfo{volume}{458}},
  \bibinfo{pages}{996} (\bibinfo{year}{2009}{\natexlab{a}}).

\bibitem[{\citenamefont{Tycko}(1983)}]{tycko83}
\bibinfo{author}{\bibfnamefont{R.}~\bibnamefont{Tycko}},
  \bibinfo{journal}{Phys. Rev. Lett.} \textbf{\bibinfo{volume}{51}},
  \bibinfo{pages}{775} (\bibinfo{year}{1983}).

\bibitem[{\citenamefont{Levitt}(1986)}]{levit86}
\bibinfo{author}{\bibfnamefont{M.~H.} \bibnamefont{Levitt}},
  \bibinfo{journal}{Prog. NMR Spect.} \textbf{\bibinfo{volume}{18}},
  \bibinfo{pages}{61} (\bibinfo{year}{1986}).

\bibitem[{\citenamefont{Cummins and Jones}(2000)}]{cummi00}
\bibinfo{author}{\bibfnamefont{H.~K.} \bibnamefont{Cummins}} \bibnamefont{and}
  \bibinfo{author}{\bibfnamefont{J.~A.} \bibnamefont{Jones}},
  \bibinfo{journal}{New J. Phys.} \textbf{\bibinfo{volume}{2}},
  \bibinfo{pages}{6} (\bibinfo{year}{2000}).

\bibitem[{\citenamefont{Cummins et~al.}(2003)\citenamefont{Cummins, Llewellyn,
  and Jones}}]{cummi03}
\bibinfo{author}{\bibfnamefont{H.~K.} \bibnamefont{Cummins}},
  \bibinfo{author}{\bibfnamefont{G.}~\bibnamefont{Llewellyn}},
  \bibnamefont{and} \bibinfo{author}{\bibfnamefont{J.~A.} \bibnamefont{Jones}},
  \bibinfo{journal}{Phys. Rev. A} \textbf{\bibinfo{volume}{67}},
  \bibinfo{pages}{042308} (\bibinfo{year}{2003}).

\bibitem[{\citenamefont{Fortunato et~al.}(2002)\citenamefont{Fortunato, Pravia,
  Boulant, Teklemariam, Havel, and Cory}}]{fortu02}
\bibinfo{author}{\bibfnamefont{E.~M.} \bibnamefont{Fortunato}},
  \bibinfo{author}{\bibfnamefont{M.~A.} \bibnamefont{Pravia}},
  \bibinfo{author}{\bibfnamefont{N.}~\bibnamefont{Boulant}},
  \bibinfo{author}{\bibfnamefont{G.}~\bibnamefont{Teklemariam}},
  \bibinfo{author}{\bibfnamefont{T.~F.} \bibnamefont{Havel}}, \bibnamefont{and}
  \bibinfo{author}{\bibfnamefont{D.~G.} \bibnamefont{Cory}},
  \bibinfo{journal}{J. Chem. Phys.} \textbf{\bibinfo{volume}{116}},
  \bibinfo{pages}{7599} (\bibinfo{year}{2002}).

\bibitem[{\citenamefont{Brown et~al.}(2004)\citenamefont{Brown, Harrow, and
  Chuang}}]{brown04}
\bibinfo{author}{\bibfnamefont{K.~R.} \bibnamefont{Brown}},
  \bibinfo{author}{\bibfnamefont{A.~W.} \bibnamefont{Harrow}},
  \bibnamefont{and} \bibinfo{author}{\bibfnamefont{I.~L.}
  \bibnamefont{Chuang}}, \bibinfo{journal}{Phys. Rev. A}
  \textbf{\bibinfo{volume}{70}}, \bibinfo{pages}{052318}
  (\bibinfo{year}{2004}).

\bibitem[{\citenamefont{M\"ott\"onen et~al.}(2006)\citenamefont{M\"ott\"onen,
  de~Sousa, Zhang, and Whaley}}]{motto06}
\bibinfo{author}{\bibfnamefont{M.}~\bibnamefont{M\"ott\"onen}},
  \bibinfo{author}{\bibfnamefont{R.}~\bibnamefont{de~Sousa}},
  \bibinfo{author}{\bibfnamefont{J.}~\bibnamefont{Zhang}}, \bibnamefont{and}
  \bibinfo{author}{\bibfnamefont{K.~B.} \bibnamefont{Whaley}},
  \bibinfo{journal}{Phys. Rev. A} \textbf{\bibinfo{volume}{73}},
  \bibinfo{pages}{022332} (\bibinfo{year}{2006}).

\bibitem[{\citenamefont{Alway and Jones}(2007)}]{alway07}
\bibinfo{author}{\bibfnamefont{W.~G.} \bibnamefont{Alway}} \bibnamefont{and}
  \bibinfo{author}{\bibfnamefont{J.~A.} \bibnamefont{Jones}},
  \bibinfo{journal}{J. Magn. Res.} \textbf{\bibinfo{volume}{189}},
  \bibinfo{pages}{114} (\bibinfo{year}{2007}).

\bibitem[{\citenamefont{Skinner et~al.}(2003)\citenamefont{Skinner, Reiss, Luy,
  Khaneja, and Glaser}}]{skinn03}
\bibinfo{author}{\bibfnamefont{T.~E.} \bibnamefont{Skinner}},
  \bibinfo{author}{\bibfnamefont{T.~O.} \bibnamefont{Reiss}},
  \bibinfo{author}{\bibfnamefont{B.}~\bibnamefont{Luy}},
  \bibinfo{author}{\bibfnamefont{N.}~\bibnamefont{Khaneja}}, \bibnamefont{and}
  \bibinfo{author}{\bibfnamefont{S.~J.} \bibnamefont{Glaser}},
  \bibinfo{journal}{J. Mag. Res.} \textbf{\bibinfo{volume}{163}},
  \bibinfo{pages}{8} (\bibinfo{year}{2003}).

\bibitem[{\citenamefont{Kobzara et~al.}(2004)\citenamefont{Kobzara, Skinner,
  Khanejac, Glaser, and Luy}}]{kobza04}
\bibinfo{author}{\bibfnamefont{K.}~\bibnamefont{Kobzara}},
  \bibinfo{author}{\bibfnamefont{T.~E.} \bibnamefont{Skinner}},
  \bibinfo{author}{\bibfnamefont{N.}~\bibnamefont{Khanejac}},
  \bibinfo{author}{\bibfnamefont{S.~J.} \bibnamefont{Glaser}},
  \bibnamefont{and} \bibinfo{author}{\bibfnamefont{B.}~\bibnamefont{Luy}},
  \bibinfo{journal}{J. Mag. Res.} \textbf{\bibinfo{volume}{170}},
  \bibinfo{pages}{236} (\bibinfo{year}{2004}).

\bibitem[{\citenamefont{Luy et~al.}(2005)\citenamefont{Luy, Kobzar, Skinner,
  Khanej, and Glaser}}]{luy05}
\bibinfo{author}{\bibfnamefont{B.}~\bibnamefont{Luy}},
  \bibinfo{author}{\bibfnamefont{K.}~\bibnamefont{Kobzar}},
  \bibinfo{author}{\bibfnamefont{T.~E.} \bibnamefont{Skinner}},
  \bibinfo{author}{\bibfnamefont{N.}~\bibnamefont{Khanej}}, \bibnamefont{and}
  \bibinfo{author}{\bibfnamefont{S.~J.} \bibnamefont{Glaser}},
  \bibinfo{journal}{J. Mag. Res.} \textbf{\bibinfo{volume}{176}},
  \bibinfo{pages}{179} (\bibinfo{year}{2005}).

\bibitem[{\citenamefont{Gershenzon et~al.}(2007)\citenamefont{Gershenzon,
  Kobzar, Luy, Glaser, and Skinner}}]{gersh07}
\bibinfo{author}{\bibfnamefont{N.~I.} \bibnamefont{Gershenzon}},
  \bibinfo{author}{\bibfnamefont{K.}~\bibnamefont{Kobzar}},
  \bibinfo{author}{\bibfnamefont{B.}~\bibnamefont{Luy}},
  \bibinfo{author}{\bibfnamefont{S.~J.} \bibnamefont{Glaser}},
  \bibnamefont{and} \bibinfo{author}{\bibfnamefont{T.~E.}
  \bibnamefont{Skinner}}, \bibinfo{journal}{J. Mag. Res.}
  \textbf{\bibinfo{volume}{188}}, \bibinfo{pages}{330} (\bibinfo{year}{2007}).

\bibitem[{\citenamefont{Gershenzon et~al.}(2008)\citenamefont{Gershenzon,
  Skinner, Brutscher, Khaneja, Nimbalkar, Luy, and Glaser}}]{gersh08}
\bibinfo{author}{\bibfnamefont{N.~I.} \bibnamefont{Gershenzon}},
  \bibinfo{author}{\bibfnamefont{T.~E.} \bibnamefont{Skinner}},
  \bibinfo{author}{\bibfnamefont{B.}~\bibnamefont{Brutscher}},
  \bibinfo{author}{\bibfnamefont{N.}~\bibnamefont{Khaneja}},
  \bibinfo{author}{\bibfnamefont{M.}~\bibnamefont{Nimbalkar}},
  \bibinfo{author}{\bibfnamefont{B.}~\bibnamefont{Luy}}, \bibnamefont{and}
  \bibinfo{author}{\bibfnamefont{S.~J.} \bibnamefont{Glaser}},
  \bibinfo{journal}{J. Mag. Res.} \textbf{\bibinfo{volume}{192}},
  \bibinfo{pages}{235} (\bibinfo{year}{2008}).

\bibitem[{\citenamefont{Sengupta and Pryadko}(2005)}]{sengu05}
\bibinfo{author}{\bibfnamefont{P.}~\bibnamefont{Sengupta}} \bibnamefont{and}
  \bibinfo{author}{\bibfnamefont{L.~P.} \bibnamefont{Pryadko}},
  \bibinfo{journal}{Phys. Rev. Lett.} \textbf{\bibinfo{volume}{95}},
  \bibinfo{pages}{037202} (\bibinfo{year}{2005}).

\bibitem[{\citenamefont{Pryadko and Quiroz}(2008)}]{pryad08a}
\bibinfo{author}{\bibfnamefont{L.~P.} \bibnamefont{Pryadko}} \bibnamefont{and}
  \bibinfo{author}{\bibfnamefont{G.}~\bibnamefont{Quiroz}},
  \bibinfo{journal}{Phys. Rev. A} \textbf{\bibinfo{volume}{77}},
  \bibinfo{pages}{012330} (\bibinfo{year}{2008}).

\bibitem[{\citenamefont{Pryadko and Sengupta}(2008)}]{pryad08b}
\bibinfo{author}{\bibfnamefont{L.~P.} \bibnamefont{Pryadko}} \bibnamefont{and}
  \bibinfo{author}{\bibfnamefont{P.}~\bibnamefont{Sengupta}},
  \bibinfo{journal}{Phys. Rev. A} \textbf{\bibinfo{volume}{78}},
  \bibinfo{pages}{032336} (\bibinfo{year}{2008}).

\bibitem[{\citenamefont{Pasini et~al.}(2008)\citenamefont{Pasini, Fischer,
  Karbach, and Uhrig}}]{pasin08a}
\bibinfo{author}{\bibfnamefont{S.}~\bibnamefont{Pasini}},
  \bibinfo{author}{\bibfnamefont{T.}~\bibnamefont{Fischer}},
  \bibinfo{author}{\bibfnamefont{P.}~\bibnamefont{Karbach}}, \bibnamefont{and}
  \bibinfo{author}{\bibfnamefont{G.~S.} \bibnamefont{Uhrig}},
  \bibinfo{journal}{Phys. Rev. A} \textbf{\bibinfo{volume}{77}},
  \bibinfo{pages}{032315} (\bibinfo{year}{2008}).

\bibitem[{\citenamefont{Karbach et~al.}(2008)\citenamefont{Karbach, Pasini, and
  Uhrig}}]{karba08}
\bibinfo{author}{\bibfnamefont{P.}~\bibnamefont{Karbach}},
  \bibinfo{author}{\bibfnamefont{S.}~\bibnamefont{Pasini}}, \bibnamefont{and}
  \bibinfo{author}{\bibfnamefont{G.~S.} \bibnamefont{Uhrig}},
  \bibinfo{journal}{Phys. Rev. A} \textbf{\bibinfo{volume}{78}},
  \bibinfo{pages}{022315} (\bibinfo{year}{2008}).

\bibitem[{\citenamefont{Pasini and Uhrig}(2008)}]{pasin08b}
\bibinfo{author}{\bibfnamefont{S.}~\bibnamefont{Pasini}} \bibnamefont{and}
  \bibinfo{author}{\bibfnamefont{G.~S.} \bibnamefont{Uhrig}},
  \bibinfo{journal}{J. Phys. A: Math. Theo.} \textbf{\bibinfo{volume}{41}},
  \bibinfo{pages}{312005} (\bibinfo{year}{2008}).

\bibitem[{\citenamefont{Biercuk
  et~al.}(2009{\natexlab{b}})\citenamefont{Biercuk, Uys, VanDevender, Shiga,
  Itano, and Bollinger}}]{bierc09b}
\bibinfo{author}{\bibfnamefont{M.~J.} \bibnamefont{Biercuk}},
  \bibinfo{author}{\bibfnamefont{H.}~\bibnamefont{Uys}},
  \bibinfo{author}{\bibfnamefont{A.~P.} \bibnamefont{VanDevender}},
  \bibinfo{author}{\bibfnamefont{N.}~\bibnamefont{Shiga}},
  \bibinfo{author}{\bibfnamefont{W.~M.} \bibnamefont{Itano}}, \bibnamefont{and}
  \bibinfo{author}{\bibfnamefont{J.~J.} \bibnamefont{Bollinger}},
  \bibinfo{journal}{Phys. Rev. A} \textbf{\bibinfo{volume}{79}},
  \bibinfo{pages}{062324} (\bibinfo{year}{2009}).

\bibitem[{\citenamefont{Uhrig and Pasini}(2009)}]{uhrig09c}
\bibinfo{author}{\bibfnamefont{G.~S.} \bibnamefont{Uhrig}} \bibnamefont{and}
  \bibinfo{author}{\bibfnamefont{S.}~\bibnamefont{Pasini}},
  \bibinfo{journal}{arXiv:0906.3605}.

\bibitem[{\citenamefont{Magnus}(1954)}]{magnu54}
\bibinfo{author}{\bibfnamefont{W.}~\bibnamefont{Magnus}},
  \bibinfo{journal}{Comm. Pure Appl. Math.} \textbf{\bibinfo{volume}{7}},
  \bibinfo{pages}{649} (\bibinfo{year}{1954}).

\bibitem[{\citenamefont{Byrd and Lidar}(2002)}]{byrd02}
\bibinfo{author}{\bibfnamefont{M.~S.} \bibnamefont{Byrd}} \bibnamefont{and}
  \bibinfo{author}{\bibfnamefont{D.~A.} \bibnamefont{Lidar}},
  \bibinfo{journal}{Quant. Inf. Proc.} \textbf{\bibinfo{volume}{1}},
  \bibinfo{pages}{19} (\bibinfo{year}{2002}).

\end{thebibliography}

\end{document}